\documentclass[twocolumn,tighten]{aastex63}

\usepackage{graphicx}
\usepackage{amsmath}
\usepackage{color}
\usepackage{ulem}
\usepackage{float}
\usepackage{threeparttable}
\usepackage{subdepth}
\usepackage{ulem}
\graphicspath{{fig/}}

\newcommand{\kms}{km~s$^{-1}$}

\newcommand{\mjybeam}{mJy~beam$^{-1}$}

\shorttitle{NOEMA observations of IRAS F07599+6508}
\shortauthors{Tan et al.}
\begin{document}

\title{Deep Observations of CO and Free-Free Emission in Ultraluminous Infrared QSO IRAS~F07599+6508}

\author[0000-0003-3032-0948]{Qing-Hua Tan}
\affil{Purple Mountain Observatory \& Key Laboratory for Radio Astronomy, Chinese Academy of Sciences, 10 Yuanhua Road, Nanjing 210023, People's Republic of China; qhtan@pmo.ac.cn}

\author[0000-0003-0007-2197]{Yu Gao} %[0000-0003-0007-2197]
\affil{Department of Astronomy, Xiamen University, 422 Siming South Road, Xiamen 361005, People's Republic of China; yugao@xmu.edu.cn}
\affil{Purple Mountain Observatory \& Key Laboratory for Radio Astronomy, Chinese Academy of Sciences, 10 Yuanhua Road, Nanjing 210023, People's Republic of China; qhtan@pmo.ac.cn}

\author[0000-0002-3331-9590]{Emanuele Daddi} %[0000-0002-3331-9590]
\affil{CEA, Irfu, DAp, AIM, Universit\`e Paris-Saclay, Universit\`e de Paris, CNRS, F-91191 Gif-sur-Yvette, France}

\author{Xiao-Yang Xia}
\affil{Tianjin Astrophysics Center, Tianjin Normal University, Tianjin 300387, People’s Republic of China}

\author[0000-0002-0901-9328]{Cai-Na Hao} %[0000-0002-0901-9328]
\affil{Tianjin Astrophysics Center, Tianjin Normal University, Tianjin 300387, People’s Republic of China}

\author[0000-0002-4721-3922]{Alain Omont} %[0000-0002-4721-3922]
\affil{Sorbonne Universit$\acute{e}$, UPMC Universit$\acute{e}$ Paris 6 and CNRS, UMR 7095, Institut d'Astrophysique de Paris, 98bis boulevard Arago, 75014 Paris, France}

\author[0000-0002-4052-2394]{Kotaro Kohno} %[0000-0002-4052-2394]
\affil{Institute of Astronomy, Graduate School of Science, The University of Tokyo, Osawa, Mitaka, Tokyo 181-0015, Japan}
\affil{Research Center for the Early Universe, Graduate School of Science, The University of Tokyo, 7-3-1 Hongo, Bunkyo-ku, Tokyo 113-0033, Japan}



\begin{abstract}
Infrared quasi-stellar objects (IR QSOs) are a rare subpopulation selected from ultraluminous infrared galaxies (ULIRGs) and have been regarded as promising candidates of ULIRG-to-optical QSO transition objects. Here we present NOEMA observations of the CO(1$-$0) line and 3~mm continuum emission in an IR QSO IRAS~F07599+6508 at $z=0.1486$, which has many properties in common with Mrk~231. The CO emission is found to be resolved with a major axis of $\sim$6.1~kpc that is larger than the size of $\sim$4.0~kpc derived for 3~mm continuum. We identify two faint CO features located at a projected distance of $\sim$11.4 and 19.1 kpc from the galaxy nucleus, respectively, both of which are found to have counterparts in the optical and radio bands and may have a merger origin. A systematic velocity gradient is found in the CO main component, suggesting that the bulk of molecular gas is likely rotationally supported. Based on the radio-to-millimeter spectral energy distribution and IR data, we estimate that about 30$\%$ of the flux at 3~mm arises from free-free emission and infer a free-free-derived star formation rate of 77 $M_\odot\ {\rm yr^{-1}}$, close to the IR estimate corrected for the AGN contribution. We find a high-velocity CO emission feature at the velocity range of about $-$1300 to $-$2000 km~s$^{-1}$. Additional deep CO observations are needed to confirm the presence of a possible very high-velocity CO extension of the OH outflow in this IR QSO.

%in the blue-shifted side of spectrum with a detection significance of $\sim 4\ \sigma$ in the integrated map. Additional deep CO observations are needed to draw a firm conclusion on the presence of a possible molecular outflow in this IR QSO.

\end{abstract}

%\keywords{editorials, notices --- 
%miscellaneous --- catalogs --- surveys}



\section{Introduction} \label{sec:intro}

Mergers of gas-rich galaxies may trigger both intense star formation and nuclear activity, leading to a violent transition from an obscured accretion and starburst stage to an optically bright quasi-stellar object (QSO) and eventually an elliptical galaxy \citep[e.g.,][]{sanders88a,sanders88b,barnes92,DiMatteo05,hopkins06}. In this merger-driven evolutionary scenario, the active galactic nucleus (AGN) was thought to be energetically dominant but obscured in the ultraluminous infrared galaxy (ULIRG; $L_{\rm IR}\geqslant 10^{12}\ L_\odot$) phase, where the most luminous sources that reach a QSO-like luminosity represent the rare ULIRG-to-QSO transition objects. These systems are very likely experiencing the so-called feedback or blowout phase that expels the gas of the host galaxy in the form of outflowing winds, which have been invoked as one of the main drivers for regulating the growth of the stellar mass and black hole (BH) accretion and producing the tight BH-bulge mass relation \citep[see][and references therein]{silk98,hopkins08,fabian12,kormendy13,king15}, and therefore represent a key phase in the lifetime of a quasar.

The large gas masses discovered in local ULIRGs via CO observations have established the presence of abundant fuel for the intense star formation activity \citep[e.g.,][]{solomon97}. Along with the clear evidence that many (U)LIRGs are strong dynamical interactions and mergers \citep{sanders96}, all the properties make these systems unique local examples of dust-enshrouded galaxies at high-$z$. \citet{xia12} have investigated for the first time the CO survey in 19 ultraluminous infrared type 1 QSOs, selected from the local ULIRGs and referred to as IR QSOs \citep[see][]{zheng02}. These IR-selected type 1 QSOs (IR QSOs) were originally compiled from the ULIRGs in the QDOT redshift survey \citep{lawrence99}, the 1 Jy ULIRG survey \citep{kim98}, and the cross-correlation of the \textit{IRAS} Point Source Catalog with the \textit{ROSAT} All-sky Survey Catalog. Studies of luminosity function for large samples of \textit{IRAS} galaxies reveal that the space densities of ULIRGs and QSOs are comparable in the local universe \citep{soifer87}, while the fraction of IR QSOs is less than 10\% of ULIRGs \citep{lawrence99}, indicating that the IR QSO phase may last only a few times 10$^7$ yr if the space density of objects is simply related to the timescale of different phases \citep{hao05}. 

The CO observations of IR QSOs show that the molecular gas masses are a few times $10^9-10^{10}\ M_\odot$ \citep{xia12}, similar to that found in local ULIRGs but significantly higher than that of optically selected PG QSOs \citep[e.g.,][]{scoville03,evans01,evans06,shangguan20}. This supports the scenario that IR QSOs are rich of fuel to feed both the star formation and the accretion of AGN. Similar conclusion has also been reached recently for type 2 QSOs, of which a sample of QSO hosts are found to be gas-rich based on the CO observations \citep{jarvis20}. Together with the far-IR excess observed in IR QSOs \citep{hao05}, and the multi-wavelengths properties \citep[i.e., large blueshifts in optical emission lines, warm IR color, intermediate slope of mid-IR spectral energy distribution (SED) between ULIRGs and PG QSOs, and high accretion rates;][]{zheng02,cao08}, all of these properties imply that IR QSOs are likely the objects caught in the short-lived transition phase between ULIRG and QSO stages \citep{canalizo01}.

Compared to the local (U)LIRGs that a large number have been spatially resolved in molecular gas emission \citep[e.g.,][]{downes98,wilson08,ueda14,xu14,xu15,imanishi19}, except Mrk~231, the nearest IR QSO at $z=0.042$ that has been well-studied in both multi-wavelength and multi-scale \citep[e.g.,][]{bryant96,taylor99,fischer10,rupke11,feruglio15,cicone20}, until recently the CO emission have been spatially resolved on sub-kiloparsec (sub-kpc) scale in a handful of IR QSOs with the ALMA \citep{tan19}. It was found that the host galaxies of IR QSOs resolved with ALMA have a diverse variety of CO morphology and kinematics (i.e., rotating disk, disturbed structure, and multiple interacting sources), indicative of more complicated evolutionary stages from merging (U)LIRGs to QSOs. This is somewhat unexpected since the objects in IR QSO phase are typically characterized by the final coalescence of the galaxies, according to the models of merger-driven evolutionary scenario \citep{barnes92,hopkins08}. However, such a classification suffers from small number statistics, spatially resolved CO observations in more IR-luminous QSOs are required for a meaningful statistical study. 

In this paper, we present CO(1$-$0) imaging of IRAS~F07599+6508 obtained with the IRAM NOrthern Extended Millimeter Array (NOEMA) interferometer. IRAS~F07599+6508 is an IR QSO at $z=0.149$ with a total IR luminosity of $3.5\times10^{12}\ L_\odot$ and AGN bolometric luminosity of $L_{\rm AGN,bol}=3.9\times10^{12}\ L_\odot$ \citep{hao05}. It is an active galaxy of particular interest because of its highly similarity to Mrk~231, the closest and most prominent template of local IR QSOs, but at a higher redshift and lack of detailed studies on the cold molecular gas of the host galaxy, especially the spatial distribution and kinematics. Similar to Mrk~231, IRAS~F07599+6508 is a low-ionization broad absorption-line (BAL) QSO with strong Fe{\sc ii} emission \citep[e.g.,][]{kwan95,veron-cetty06}, which is a very rare subclass of BAL QSOs that may represent young, heavily enshrouded AGNs, where the cocoon of gas and dust has a large covering factor \citep{boroson92,zheng02}.

The IRAM 30m CO observations show massive amounts of molecular gas residing in IRAS~F07599+6508 and a broad FWHM line width of $380\pm14$ \kms\ and $490\pm40$ \kms\ for the CO(1$-$0) and CO(2$-$1) line, respectively \citep{xia12}. Based on six independent methods, \citet{veilleux09} estimate an average fraction of AGN contribution to the bolometric luminosity of $\alpha_{\rm AGN}=0.75$, consistent with other IR QSOs (e.g., $\sim 70\%$ for Mrk~231) where the $\alpha_{\rm AGN}$ is found to be typically higher than that ($\sim 35\%-40\%$) of local cooler ULIRGs. The weakness of both soft ($L_{\rm 0.1-2.4\ keV}=6.9\times10^{42}\ {\rm erg\ s^{-1}}$) and hard X-ray emission ($L_{\rm 2-10\ keV}=1.3\times10^{42}\ {\rm erg\ s^{-1}}$) seen in IRAS~F07599+6508 \citep[e.g.,][]{xia01,brightman11,luo14,la-caria19}, i.e., the X-ray emission is not at the level expected from its optical/UV emission compared to typical QSOs, imply that its central AGN is very likely highly obscured. %%%%%%%{\bf Analysis of $ROAST$ soft X-ray data show a similar absorbing column density $N_{\rm H}$ of about $4\times10^{20}$ cm$^{-2}$ for both IRAS~F07599+6508 and Mrk~231 \citep{xia01}.} 

The existence of multi-phase outflow winds in IRAS~F07599+6508 has been demonstrated by the detection of large blueshifts ($\sim$2000 \kms) of optical emission lines, Na~{\sc i}~D interstellar absorption line, and OH 119$\mu$m P-Cygni/blue-shifted absorption profile, closely resemble to Mrk~231 \citep{boroson92,zheng02,spoon13,veilleux13,rupke17}. The sensitive CO(1$-$0) mapping observations will help constrain the distribution and kinematics of the molecular gas reservoir in IRAS~F07599+6508 and offer the possibility of detecting the molecular outflow based on the presence of broad wings in the CO line emission. Tremendous amount of observing time were devoted to Mrk~231 on NOEMA revealing the outstanding molecular outflows including in dense gas tracers \citep{aalto12}, and likely the tentative detection of large-scale molecular oxygen emission \citep{wang20}. 

Alike to Mrk~231, IRAS~F07599+6508 has typical features expected for a system transiting from the heavily obscured starburst phase to the unobscured QSO phase, and could also be regarded as a representative of IR QSOs which have low number density in the local universe but are expected to be common at high redshift \citep[][]{veilleux99,lefloch05}, and thus provides nearby laboratory to probe the ULIRG-to-optical QSO transition phase that may hold the key to understanding the galaxy evolution in a merger-driven scenario.

The paper is organized as follows. In Section~\ref{sec:obs}, we describe the NOEMA observations and data reduction process. Section~\ref{sec:results} presents the results from the analysis of 3~mm continuum and CO(1$-$0) line data, including morphology, spatially decomposed CO components, kinematics, and an analysis of archival radio continuum data from VLA. In section~\ref{sec:diss}, we discuss the nature of 3~mm continuum emission, the origin of the off-center CO components, and make a comparison of the molecular gas properties between IRAS~F07599+6508 and Mrk~231. We summarize our main findings in Section~\ref{sec:sum}. Throughout this work, we assume a standard $\Lambda$CDM cosmology with $H_{\rm 0}$ = 70 \kms\ Mpc$^{-1}$, $\Omega_{\rm m}$ = 0.3, and $\Omega_\Lambda$ = 0.7.

\section{Observations and data reduction} \label{sec:obs}

\subsection{NOEMA CO(1-0) Observations} \label{subsec:obsnoema}

We have observed the CO(1$-$0) line and the continuum emission in IRAS F07599+6508 with the IRAM NOEMA interferometer between May and July 2018. The observations were obtained with a total on-source time of 12.7 h in the compact array D-configuration (with eight antennas covering baselines of 15.2 - 175.9 m). We used the PolyFix correlator in 3 mm Band 1, which provides 2$\times$8 GHz of instantaneous dual-polarisation bandwidth with a spectral resolution of 2 MHz (corresponding to 6.0 \kms at the redshifted CO(1$-$0) frequency of 100.353 GHz). The tuning frequency was set to 98.2 GHz with frequency ranged from 94.2 to 102.2 GHz in the Upper SideBand (USB) and 78.8 to 86.8 GHz in the Lower SideBand (LSB). In this paper, we analyze the CO(1-0) line and the continuum data in the USB; the rest of detected lines (i.e., CN(1-0) lines at 113.191 and 113.491 GHz, CO(1$-$0) isotopologue lines, and the 109.174 GHz HC$_3$N (12$-$11) line) covered in the USB and the LSB line data will be presented in a future paper.

The data calibration was performed using the GILDAS\footnote{\url{http://iram.fr/IRAMFR/GILDAS}; for more information, see \cite{guilloteau00}} package CLIC with help from the staff in Grenoble. The typical uncertainty of flux calibration in the 3 mm band is about 10\%. We binned the $uv$ tables by 20 \kms and redefined the rest frequency to the sky frequency of CO(1$-$0) transition. We then used the software GILDAS/MAPPING for image cleaning and analysis in the $uv$-plane. We also exported the data to a CASA\footnote{\url{http://casa.nrao.edu}; see \citet{mcmullin07}}-compatible format for a comparison analysis. The continuum $uv$ table was created from the calibrated visibilities by averaging the line-free channels using the MAPPING task \texttt{uv\_continuum} with line emission channels filtered (\texttt{uv\_filter} task). The continuum emission was subtracted from the line $uv$-data by fitting a first-order polynomial baseline for each visibility (\texttt{uv\_baseline} task). The USB and LSB continuum $uv$-data were combined using the \texttt{uv\_merge} task to form a single output table centered at 90.5 GHz. The synthesized beam size using natural weighting is 5.\arcsec67$\times$3.\arcsec81 and the NOEMA primary beam at the sky frequency of CO(1$-$0) is 50.\arcsec2$\times$50.\arcsec2. We obtain a sensitivity of $\sim$0.44 \mjybeam\ in the spectral velocity resolution of 20 \kms\ and $\sim$10~$\mu$Jy beam$^{-1}$ for the continuum image.

\subsection{Archival VLA Radio Data} \label{subsec:obsvla}

In this work we also make use of ancillary radio continuum data observed with the Very Large Array (VLA) for a comprehensive analysis. IRAS~F07599+6508 has been observed at 1.49, 4.86, 8.44, 14.94, and 22.46 GHz in CD configuration (project ID: AB0783). Higher resolution observations in AB configuration (project ID: AH0406) were available at 1.49, 4.86, and 8.44 GH, as well as in A configuration at 14.94 GHz (project ID: AN0104). We obtained the raw data sets from the VLA archive\footnote{\url{https://science.nrao.edu/facilities/vla/archive/index}}. The data flagging and reduction were performed by using the AIPS package \citep{vanmoorsel96}. The full image at each frequency and array was processed individually with the AIPS CLEAN task \texttt{imagr} and then transferred to FITS format for further analysis with the CASA package. The synthesized beam using Briggs weighting with a robustness parameter of 0.5 ranges from 65.\arcsec2$\times$28.\arcsec3 at 1.49 GHz to 1.\arcsec67$\times$1.\arcsec30 at 43.34 GHz for observations in CD configuration, and from 5.\arcsec33$\times$2.\arcsec08 at 1.49 GHz to 0.\arcsec18$\times$0.\arcsec13 at 14.94 GHz for higher resolution observations in AB and A configurations.

\section{Results and analysis} \label{sec:results}

\subsection{3~mm Continuum Emission} \label{subsec:cont}

%%%%%%%%%%%%%%%- Fig-1 -%%%%%%%%%%%%%%%%
\begin{figure*}[tbp!]
\centering
\includegraphics[width=0.26\linewidth]{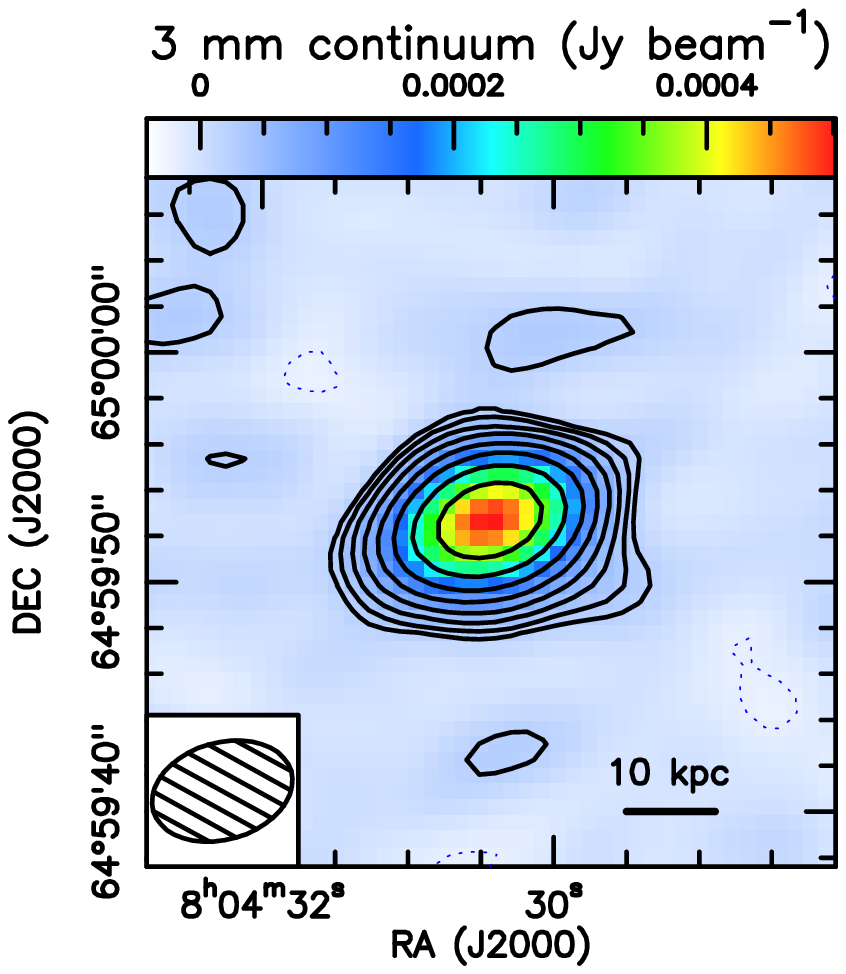}
\hspace{8pt}
\includegraphics[width=0.275\linewidth]{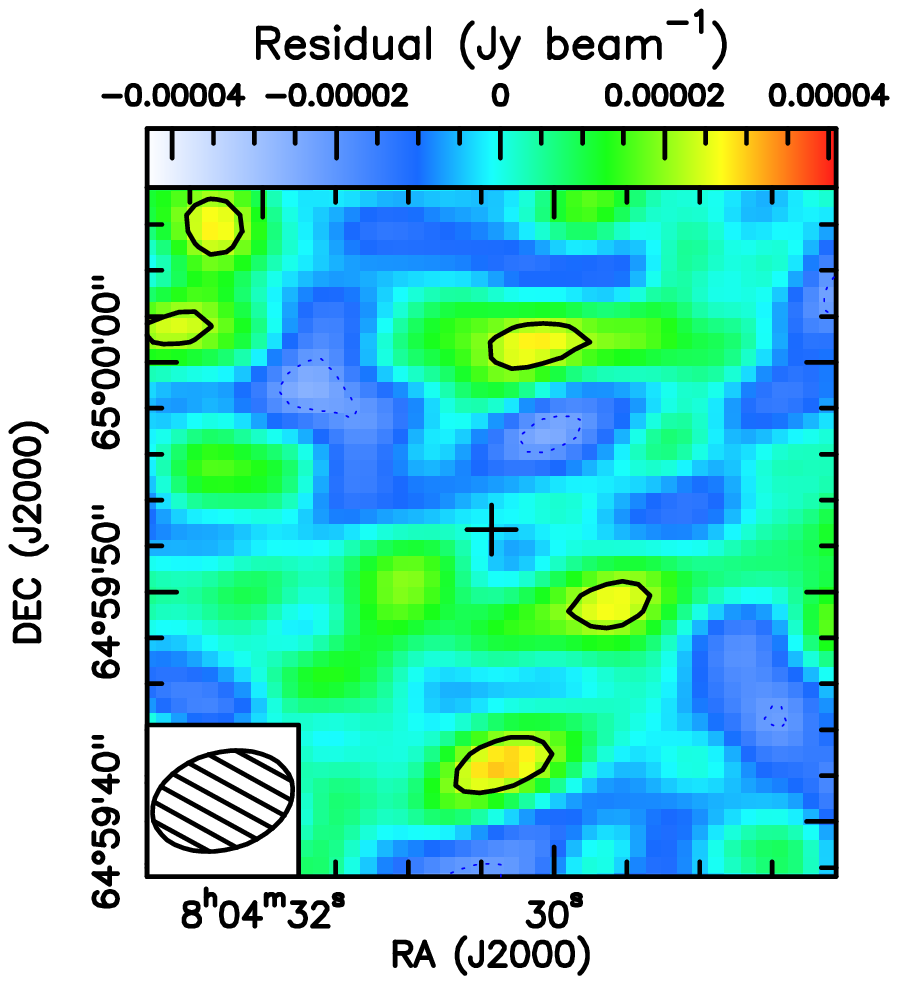}
\includegraphics[width=0.285\linewidth]{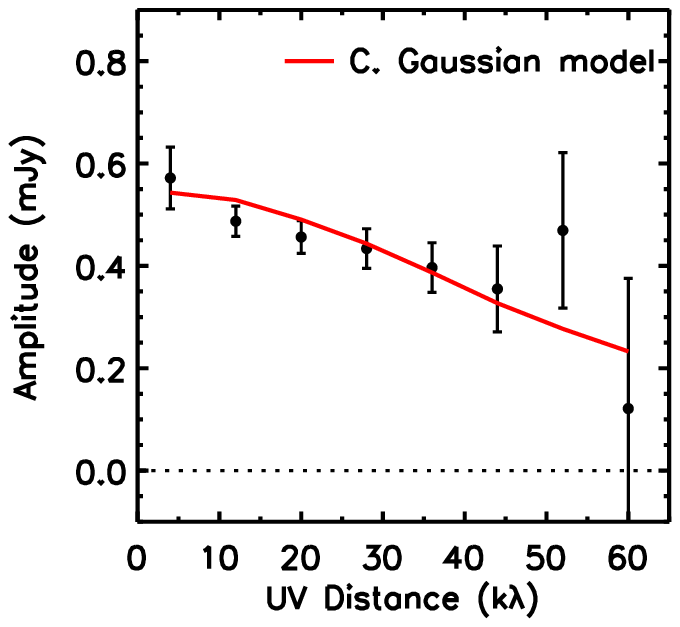}
\caption{NOEMA cleaned map of 3~mm continuum emission (left), the residual map of 3~mm emission after subtracting a circular Gaussian model (middle), and the corresponding $uv$-plot showing the visibility amplitude as a function of the $uv$-distance (right) for IRAS~F07599+6508. For 3~mm continuum map, contours start at $\pm 2\sigma$ (1$\sigma$ rms noise is 10 $\mu$Jy beam$^{-1}$) and increase by a factor of 1.5, with positive (negative) contours shown as solid (dotted) lines, while for the residual map, contours start at $\pm 2\sigma$ and increase in steps of 1$\sigma$. The synthesized beam of 6.\arcsec25$\times$4.\arcsec22 with position angle 106.2$^\circ$ is shown in the bottom left corner. The plus sign indicates the centroid of the 3 mm continuum emission derived from a circular Gaussian fit to the $uv$-data. In the right panel, the black circles with error bars shown in the $uv$-plot are the data binned in $uv$-radius steps of 8 k$\lambda$. Error bars show the statistical photon noise on the average amplitudein each bin. The red curve indicates the best-fitting model with a circular Gaussian function to the $uv$-data. \label{fig:cont} }
\end{figure*}
%%%%%%%%%%%%%%%%%%%%%%%%%%%%%%%%%%%%%%%%

The continuum emission at 90.5 GHz (rest-frame 103.9 GHz; $\sim$3~mm) is clearly detected towards the nucleus of IRAS~F07599+6508. Figure~\ref{fig:cont} shows the 3~mm continuum map and the corresponding $uv$-plot that shows the amplitude of the visibilities as a function of the $uv$-distance. It can be seen from the $uv$-data that the 3~mm continuum source is resolved in our observations, as an unresolved source would give a flat distribution of visibilities. 

Assuming that the surface brightness distribution of the continuum follow a symmetric Gaussian profile, we performed a fit to the visibilities in GILDAS using task \texttt{uv\_fit} and found that the continuum source is best fitted by a two-dimensional (2D) circular Gaussian model with a deconvolved full width at half maximum (FWHM) size of 1.\arcsec53$\pm$0.\arcsec19 ($\sim 4.0\pm0.5$ kpc) and a flux density of 560$\pm$20 $\mu$Jy (see Table~\ref{tab:results}). A 2D elliptical Gaussian model fit to the continuum visibilities returns a deconvolved FWHM size of (1.\arcsec54$\pm$0.\arcsec22)$\times$(1.\arcsec50$\pm$0.\arcsec39) and a flux density of 560$\pm$20 $\mu$Jy, which are in good agreement with those derived from fit with a circular Gaussian model, but with slightly larger uncertainties on the size measurement. In addition, the position angle ($0^\circ\pm293^\circ$) derived from elliptical Gaussian fit is rather uncertain due to the large uncertainty. 

In the middle panel of Figure~\ref{fig:cont}, we show the residual after removing the fit with a circular Gaussian model. No significant peaks or negative regions with signal-to-noise ratio (SNR) $\gtrsim 3$ are found in the residual map. The red curve in $uv$-plot (see the right panel of Figure~\ref{fig:cont}) shows the result of circular Gaussian model fit. Moreover, it can be seen from the right panel of Figure~\ref{fig:cont}, most data points are consistent with the model fit (see the red curve) with a circular Gaussian function within 1$\sigma$. All of these suggest that a single circular Gaussian model may provide a good description to the 3~mm continuum emission based on the current data.

As an independent check on the results from our analysis of $uv$-data, we also measured the size and flux of the continuum emission by fitting the data in the image plane using the \texttt{imfit} task in CASA. A 2D Gaussian fit to the 3~mm continuum map of Figure~\ref{fig:cont} finds an deconvolved FWHM size of (1.\arcsec84$\pm$0.\arcsec45)$\times$(1.\arcsec40$\pm$0.\arcsec49) and an integrated flux of 550$\pm$20 $\mu$Jy, consistent with the estimates from the \texttt{uv\_fit} within 1$\sigma$ uncertainties.

%%%%%%%%%%%%%%%- Fig-2 -%%%%%%%%%%%%%%%%
\begin{figure*}[htbp]
\centering
\includegraphics[width=0.43\linewidth]{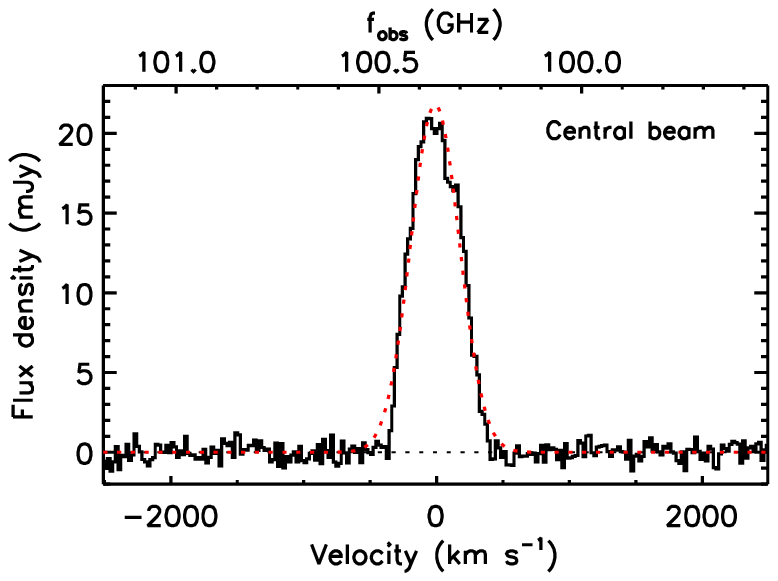}
\hspace{8pt}
\includegraphics[width=0.27\linewidth]{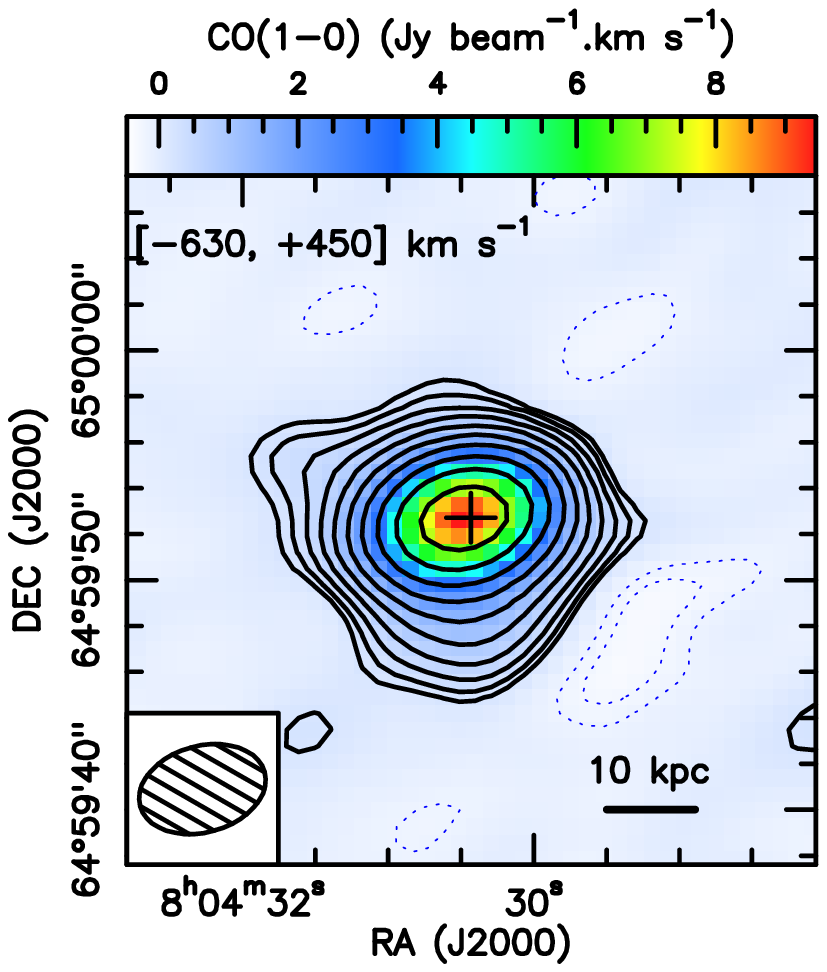}
\caption{NOEMA continuum-subtracted spectrum (left) and cleaned map (right) of CO(1$-$0) emission of IRAS~F07599+6508. The CO(1$-$0) spectrum is extracted from the central beam and the spectrum is binned to a channel width of $\sim$20 \kms. The red dotted line represents a Gaussian fit to the spectrum. The zero velocity corresponds to the redshift of $z_{\rm CO}=0.1486$ derived from a Gaussian fit to the spectrum. The map of CO(1$-$0) emission is integrated in the [$-$630, +450] \kms\ velocity range. Contours start at $\pm 2\sigma$ (1$\sigma$ rms noise is 0.092 Jy beam$^{-1}$ km s$^{-1}$) and increase by a factor of 1.5, with positive (negative) contours shown as solid (dotted) lines. The synthesized beam (5.\arcsec61$\times$3.\arcsec77) is shown in the bottom left corner of the map. The plus sign indicates the peak of 3 mm continuum emission. \label{fig:co} }
\end{figure*}
%%%%%%%%%%%%%%%%%%%%%%%%%%%%%%%%%%%%%%%%

\subsection{CO(1$-$0) Line Emission}\label{subsec:co}

\subsubsection{CO morphology and spatial decomposition} \label{subsubsec:decomp}

Figure~\ref{fig:co} shows the continuum-subtracted CO(1$-$0) spectrum and the integrated intensity map of IRAS~F07599+6508. The CO spectrum is extracted by fitting a point-source model to the $uv$-table using the \texttt{uv\_fit} task. By fitting a Gaussian profile to the spectrum, we measure a redshift of $z_{\rm CO}=0.1486$, being used to set the systematic velocity. The velocity width (FWHM) derived from the Gaussian fit to the spectrum is 430$\pm$10 \kms. We note that the CO(1$-$0) line profile is asymmetrical and a single Gaussian function does not provide a good fit to the line peak and wings. A double-peak profile is detected at the line center albeit not very prominent, which is often taken to indicate orbital motion in a merger or a rotating disk \citep[e.g.,][]{downes98,greve05,daddi10}. In addition, the line profile also shows a slightly weaker peak at velocity $\sim 100-150$ \kms, indicating that the motion of molecular gas in the host galaxy is complex.

The irregular morphology of the CO emission with extended structure is shown by the map in the right panel of Figure~\ref{fig:co}, obtained by collapsing the channels in the [$-$630, +450] \kms\ velocity range where the flux is positive in each channel when the CO spectrum is rebinned to a channel width of $\sim 60$~\kms. The CO emission appears to be resolved with an extension to the northeast and to the south of the nucleus, respectively. To characterize the distribution of CO emission, we fitted an elliptical Gaussian model to the $uv$-data with channels averaged in the [$-$630, +450] \kms\ velocity range using the task \texttt{uv\_fit}. The best fit gives a source size FWHM of $(2.\arcsec 33 \pm 0.\arcsec11)\times(1.\arcsec 91 \pm 0.\arcsec07)$. The peak of CO emission is found to be co-spatial with the 3~mm continuum emission. 

\begin{figure*}[htbp]
\centering
\includegraphics[width=0.3\linewidth]{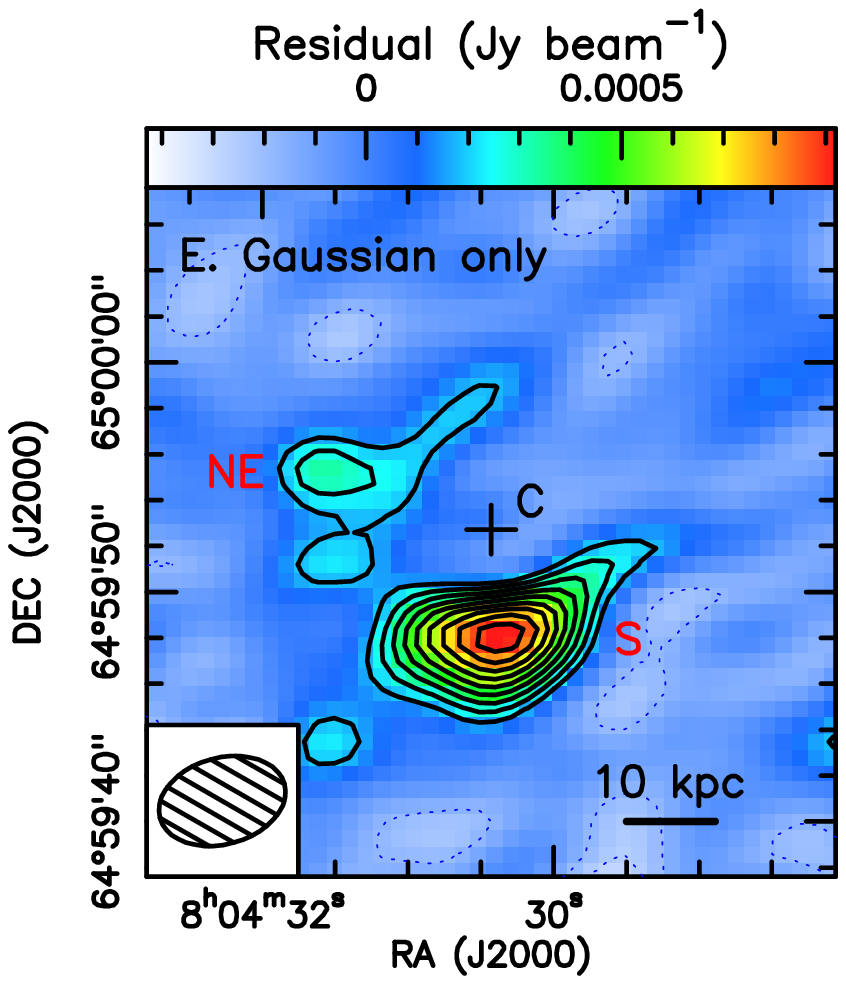}
\hspace{8pt}
\includegraphics[width=0.3\linewidth]{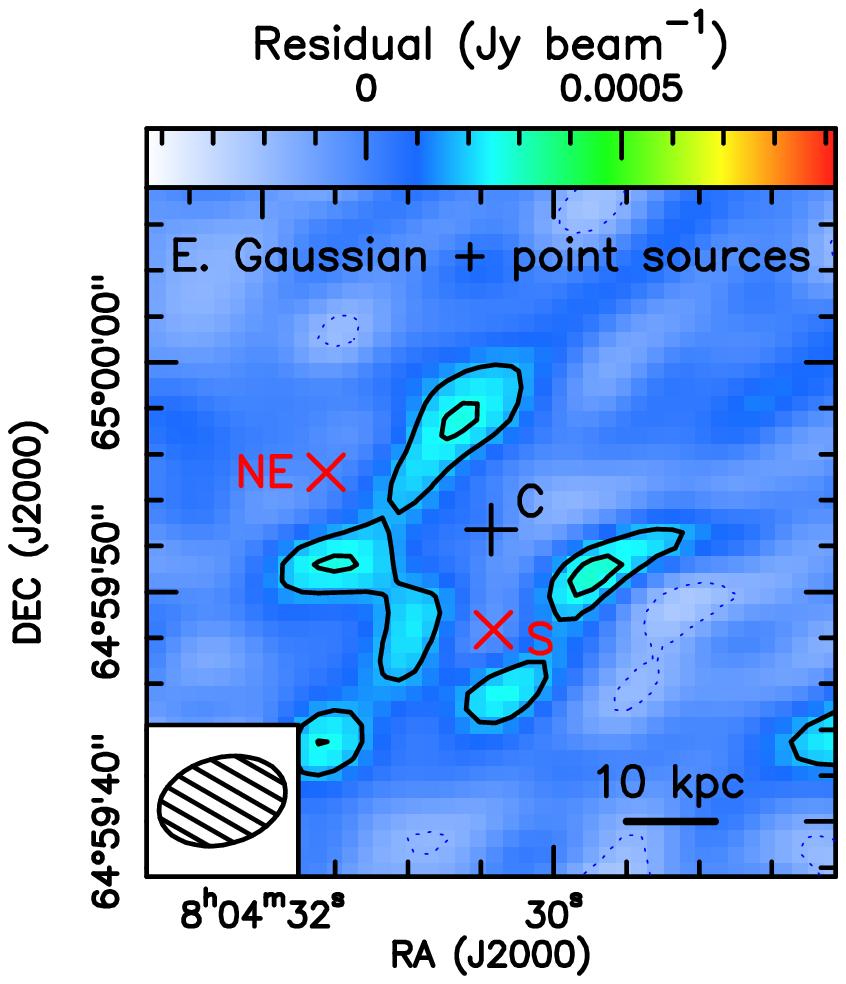}
\caption{Residual maps of CO(1$-$0) emission after subtracting a single elliptical Gaussian model (left) and a combination of an ellipical Gaussian model and two point sources models (right). The black plus sign marks the peak of 3~mm continuum emission, while the two red crosses indicate the peak positions of the two point-like CO feature. Contours start at $\pm 2\sigma$ and increase in steps of 1$\sigma$ for both maps. The two point-like subcomponents are labeled as F07599S and F07599NE, respectively.\label{fig:residuals} }
\end{figure*}

In the residual map after subtraction of an elliptical Gaussian component, we detect two off-center CO subcomponents at above 5$\sigma$, which we denote as northeast (NE) and south (S), and we refer to the main component fitted with an elliptical Gaussian in the galaxy center as F07599C (see Figure~\ref{fig:residuals}). To take into account the off-center components, we remake the fitting in the $uv$-plane using combinations of functions by allowing all parameters to be free in the modeling, and the details of the model fitting are described in Appendix~\ref{appdix:decomp}. The best fit shows that the CO emission can be spatially decomposed into three components, an elliptical Gaussian source (F07599C) and two point sources (F07599S and F07599NE).

For the main component F07599C, the model fit returns a source size FWHM of $(2.\arcsec 35 \pm 0.\arcsec11)\times(1.\arcsec 66 \pm 0.\arcsec09)$, corresponding to 6.1 by 4.3 kpc, which is about two times of the median source size ($\sim$3.2 kpc) measured for the local IR QSOs with  well-resolved CO morphology \citep{tan19}. The three components identified in the CO emission and their parameters are listed in Table~\ref{tab:results}.

%%%%%%%%%%%%%%%- Fig-3 -%%%%%%%%%%%%%%%%
%%%%%%%%%%%%%%%%%%%%%%%%%%%%%%%%%%%%%%%%

%%%%%%%%%%%%%%- Table-1- %%%%%%%%%%%%%%

\begin{deluxetable*}{lccccccc}[htbp]
\tablenum{1}
\centering
\tabletypesize{\scriptsize}
\addtolength{\tabcolsep}{-1.pt}
\tablecaption{Measured properties of CO(1$-$0) line and 3~mm continuum emission of IRAS~F07599+6508}\label{tab:results}
\tablewidth{0pt}
\tablehead{
\colhead{Emission} & \colhead{Component} & \colhead{R.A.} & \colhead{Decl.} & \colhead{Size FWHM}  & \colhead{FWHM$_{\rm CO(1-0)}$} & \colhead{$v_{\rm offset}$} & \colhead{Integrated Flux} \\
    & & \colhead{(J2000)} & \colhead{(J2000)} & \colhead{(arcsec)}   & \colhead{(\kms)} & \colhead{(\kms)} &   \\
    \colhead{(1)} &  \colhead{(2)} & \colhead{(3)} & \colhead{(4)} & \colhead{(5)} & \colhead{(6)} & \colhead{(7)} & \colhead{(8)} 
    }
\startdata
CO(1$-$0) & F07599C & 08:04:30.468$\pm$0.004 & 64:59:52.72$\pm$0.02 & (2.35$\pm$0.11)$\times$(1.66$\pm$0.09) & 430$\pm$10 & 0 & 11.24$\pm$0.09 Jy \kms\\
          & F07599S & 08:04:30.41$\pm$0.03 & 64:59:48.36$\pm$0.12 & \nodata & 100$\pm$10 (narrow) & 100$\pm$10 & 1.12$\pm$0.07 Jy \kms \\
          &  &  &  &  &  500$\pm$80 (broad) &  & \\
          & F07599NE & 08:04:31.56$\pm$0.07 & 64:59:55.2$\pm$0.3 & \nodata & 390$\pm$70 & 40$\pm$30 & 0.45$\pm$0.07 Jy \kms \\
\hline
3~mm & total & 08:04:30.432$\pm$0.010 & 64:59:52.71$\pm$0.04 & 1.53$\pm$0.19 & & & 560$\pm$20 $\mu$Jy \\
\enddata
\tablecomments{Column (1): CO(1$-$0) line and 3~mm continuum emission. Column (2): components of CO(1$-$0) emission from source spatial decomposition in the $uv$-plane. Column (3) and (4): source position derived from model fit to the $uv$-data. Column (5): FWHM source size derived from model fitting to the visibilities using the GILDAS task \texttt{uv\_fit}. Dots in the column indicate an unresolved component. The physical scale of 1\arcsec corresponds to $\sim$2.6 kpc. Column (6): FWHM CO line width derived from a Gaussian profile fit to the spectrum (except for F07599S, where the CO spectrum is fitted simultaneously with two Gaussians consist of one narrow and one broad components centered at the same velocity). Column(7): CO(1$-$0) line centroid velocity offset from the systematic velocity. The centroid velocity is derived from a Gaussian fit to the spectrum. Column (8): CO(1$-$0) velocity-integrated intensity and 3~mm continuum emission flux. The CO integrated fluxes are measured from the spectra shown in Figure~\ref{fig:3comps} by integrating the intensity over the line-emitting region in each channel. The continuum flux is measured from a circular Gaussian fit to the $uv$-data. Errors in fluxes quoted in this table are statistical and do not include the absolute flux calibration uncertainty.}
%\tablenotetext{a}{}
\end{deluxetable*}
%%%%%%%%%%%%%%%%%%%%%%%%%%%%%%%%%%%%%%

%%%%%%%%%%%%%%%- Fig-4 -%%%%%%%%%%%%%%%%
\begin{figure*}[tbp!]
\centering
\includegraphics[width=0.7\linewidth]{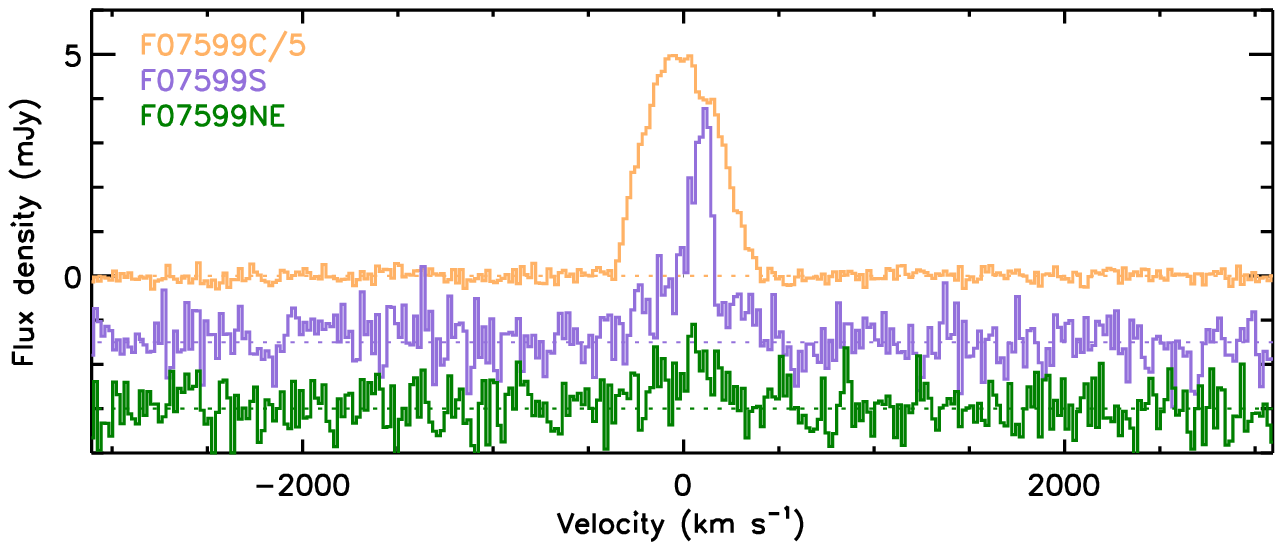}
%\hspace{-12pt}
\includegraphics[width=0.3\linewidth]{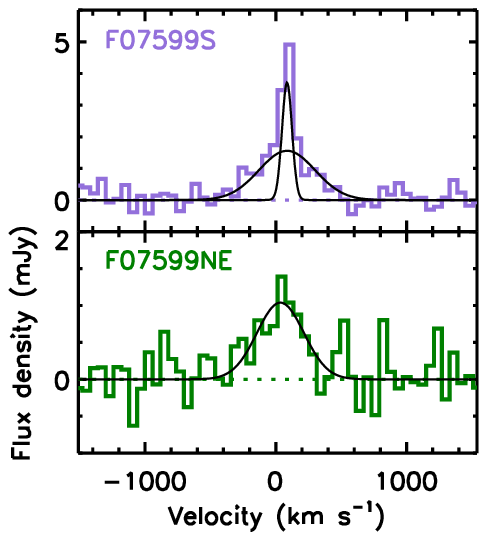}
\hspace{5pt}
\includegraphics[width=0.385\linewidth]{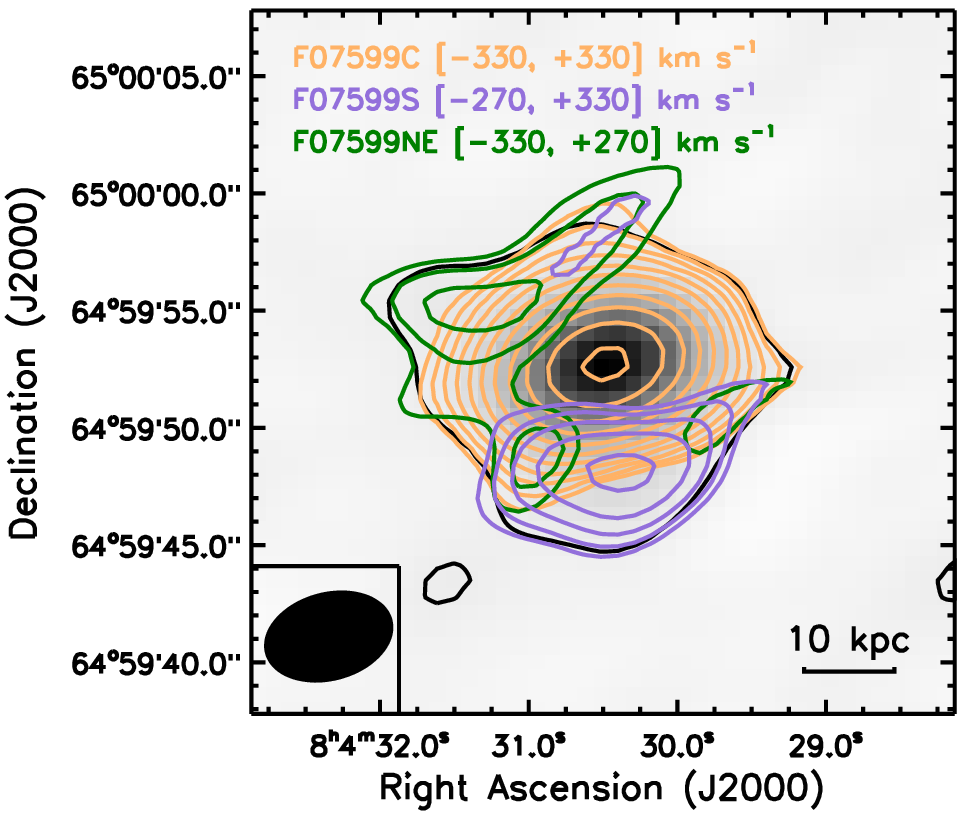}
\caption{Top: CO(1$-$0) spectra of the three components extracted from GILDAS task \texttt{uv\_fit} using an elliptical Gaussian and two point-source functions for F07599C (orange), F07599S (purple) and F07599NE (green), respectively. The CO spectrum of F07599C is divided by 5 for comparison purpose. All spectra are smoothed to a velocity resolution of $\sim$20~\kms.  Bottom-left: zoomed-in views of CO(1$-$0) spectra for F07599S and F07599NE and the spectra are re-binned in channels of $\sim$60~\kms. The black lines show the Gaussian fits to the spectra. For F07599S, we modeled the line profile with two Gaussians (one narrow and one broad), while for F07599NE the spectrum is fitted by using a single Gaussian. Bottom-right: the corresponding continuum-subtracted CO(1$-$0) maps (colored contours) of the three components overlaid on the CO(1$-$0) map of the whole IRAS~F07599+6508 (greyscale; same as the map in the right panel of Figure~\ref{fig:co}). The colored contours show the CO(1$-$0) emission integrated in the [-330, +330] \kms\ (orange), [-270, +330] \kms\ (purple), and [-330, +270] \kms\ (green) velocity range, respectively, while the black contour indicates the global CO(1$-$0) emission detected at a 3$\sigma$ significance in the intensity map integrated in the [$-$630, +450] \kms\ velocity range. The colored contours start at $2\sigma$ and increase by a factor of 1.5. The 1$\sigma$ rms noise is 0.075, 0.095, and 0.082 Jy beam$^{-1}$ km s$^{-1}$ for F07599C, F07599S, and F07599NE, respectively. The CO map of F07599C was derived from a cleaned data cube, while the maps of F07599S and F07599NE were not cleaned given the relatively low SNR.  \label{fig:3comps} }
\end{figure*}
%%%%%%%%%%%%%%%%%%%%%%%%%%%%%%%%%%%%%%%%

\subsubsection{Properties of the decomposed CO components} \label{subsubsec:subcomp}

Figure~\ref{fig:3comps} shows the CO(1$-$0) spectra of the three components and the corresponding velocity-integrated CO maps. The CO spectra of the three components are extracted simultaneously from the line cube in the $uv$-plane with fitting models identical to the best-fit models summarized in Table~\ref{tab:results} (see Section~\ref{subsubsec:decomp}). The spectrum of F07599C is obtained with $uv$-modeling fit using an elliptical Gaussian function by fixing the parameters of centroid and major/minor axes, while for the off-center components, F07599S and F07599NE, which are offset from the nucleus of main component with a projected distance of 4.\arcsec 4 ($\sim$ 11.4 kpc) and 7.\arcsec 4 ($\sim$ 19.1 kpc), respectively, the spectra are extracted by a point source model with centroids fixed. 

A single Gaussian fit to the line profile of F07599C and F07599NE shown in Figure~\ref{fig:3comps} yields a peak flux density of 26.0$\pm$0.3 mJy and 1.0$\pm$0.2 mJy, and an FWHM of 430$\pm$10 \kms\ and 390$\pm$70 \kms, respectively. F07599NE has a velocity offset of 40$\pm$30 \kms , consistent within the uncertainties with the systematic velocity. For F07599S, the spectrum is best fitted by two nested Gaussians (with a shared centre), e.g., a narrow Gaussian to fit the central core of the line, and a broad Gaussian to fit the broad emission. The best-fit gives a peak flux density of 3.9$\pm$0.4 mJy and 1.3$\pm$0.3 mJy, and an FWHM of 100$\pm$10 \kms\ and 500$\pm$80 \kms\ with velocity centroid redshifted at a velocity of about 100$\pm$10 \kms\ for the narrow and broad emission, respectively. The results of Gaussian fits derived for the three components are reported in Table~\ref{tab:results}. 

Similar to the CO spectrum extracted from the central beam, the CO line profile of F07599C shows characteristic double peak at the line center and a weaker peak at the velocity $\sim 100-150$ \kms, which is found to be coincident with the CO line centroid velocity of F07599S (see Figure~\ref{fig:3comps}, top panel). The velocity coincidence is likely to suggest that a significant portion of molecular gas are detected in the overlap region between F07599C and F07599S, where the perturbation of molecular gas make an imprint on the CO line profile. Higher spatial resolution observations are needed to resolve the emission and clarify the nature of the source.

The CO morphology of each component is shown by the map in the bottom-right panel of Figure~\ref{fig:3comps}, obtained by averaging the velocity range of [$-$330, +330] \kms, [$-$270, +330] \kms, and [$-$330, +270] \kms, where the emission is detected in consecutive channels at above 1$\sigma$, for F07599C, F07599S, and F07599NE, respectively. The velocity-integrated CO emission of each component was imaged separately by collapsing the velocity channels over the line-emitting region. It can be seen from Figure~\ref{fig:3comps} that the superimposed image of the three components agrees well with the global CO map that is integrated in the [$-$630, +450] \kms\ velocity range with coincident image contours (e.g., contours at a 3$\sigma$ level). We derive an integrated flux density of 11.24$\pm$0.09 Jy \kms , 1.12$\pm$0.07 Jy \kms , and 0.45$\pm$0.07 Jy \kms , by integrating the intensity over the line-emitting region in each channel of the spectra shown in Figure~\ref{fig:3comps}, for F07599C, F07599S, and F07599NE, respectively. From this analysis, the two subcomponents (F07599S and F07599NE) are identified at 16.0$\sigma$ and 6.4$\sigma$ in the intensity maps of residual data cube, respectively. These weak features are also clearly present in the channel maps, where extended structures are found in the S and NE direction (see Appendix~\ref{appdix:channel}).

%%%%%%%%%%%%%%%- Fig-5 -%%%%%%%%%%%%%%%%
\begin{figure*}[htbp]
\centering
\begin{minipage}[b]{0.43\textwidth}
\centering
\includegraphics[width=0.33\linewidth]{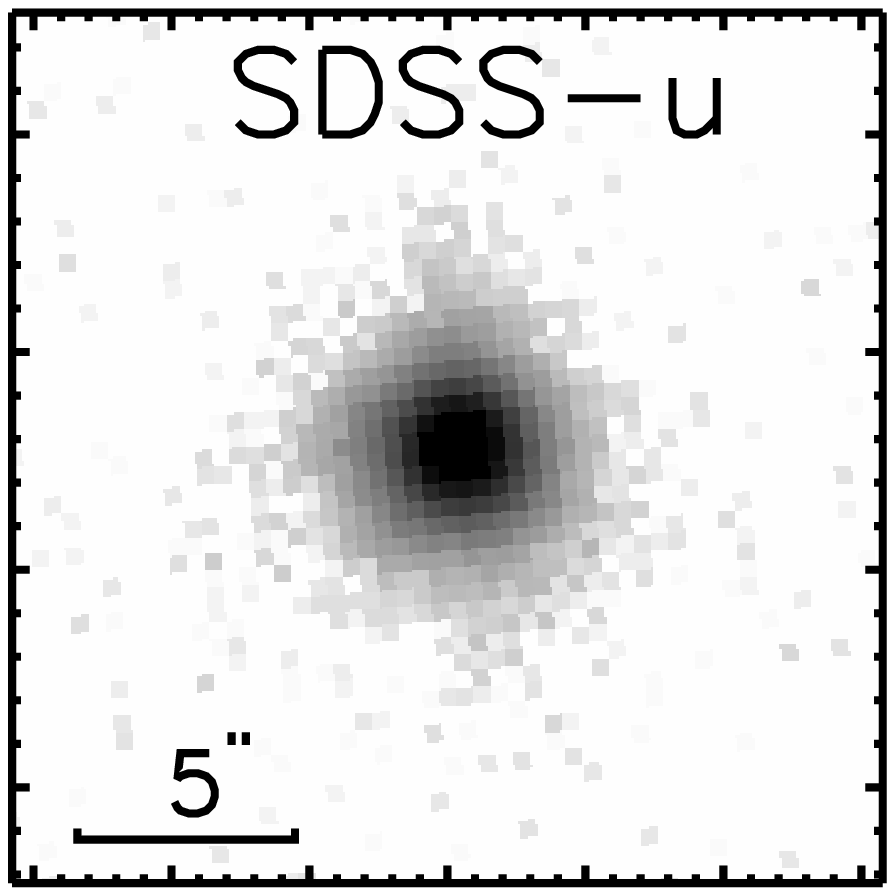}
\hspace{-8pt}
\includegraphics[width=0.33\linewidth]{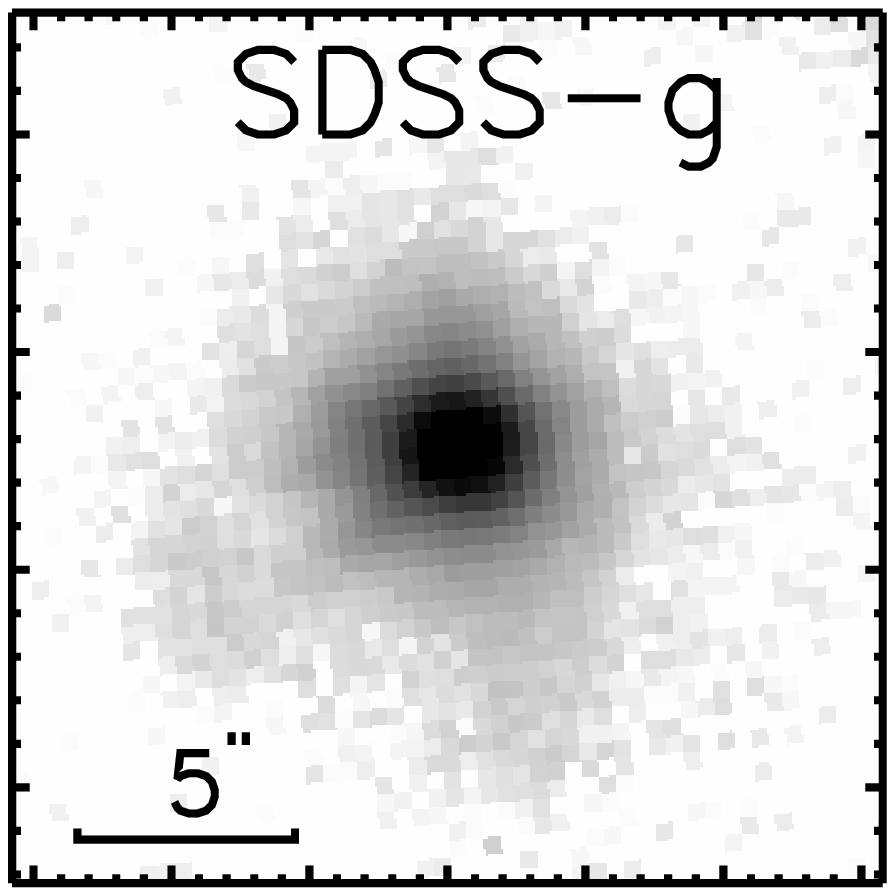}
\hspace{-8pt}
\includegraphics[width=0.33\linewidth]{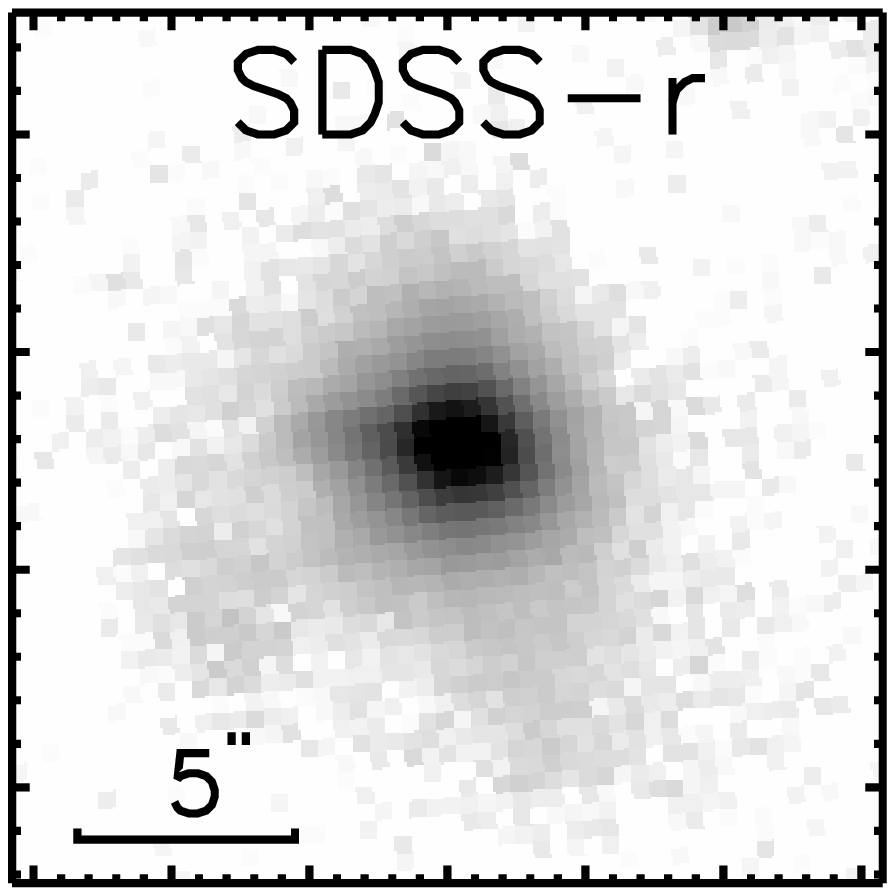}
%\vspace{-8pt}
%\hspace{-7pt}
\includegraphics[width=0.33\linewidth]{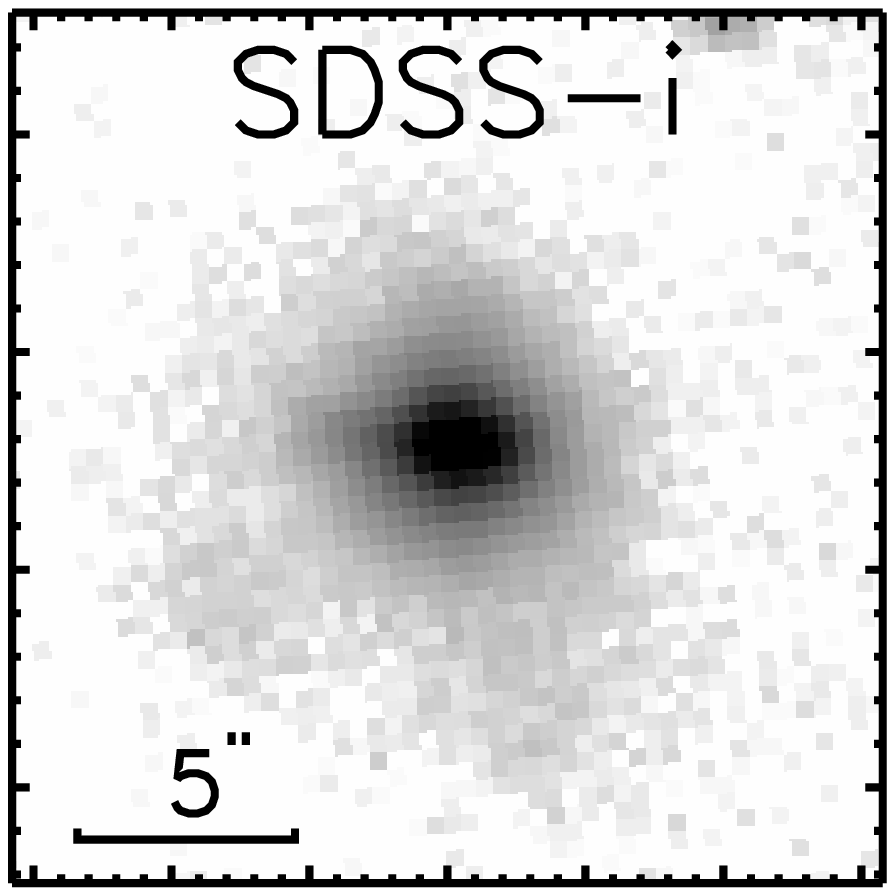}
\hspace{-8pt}
\includegraphics[width=0.33\linewidth]{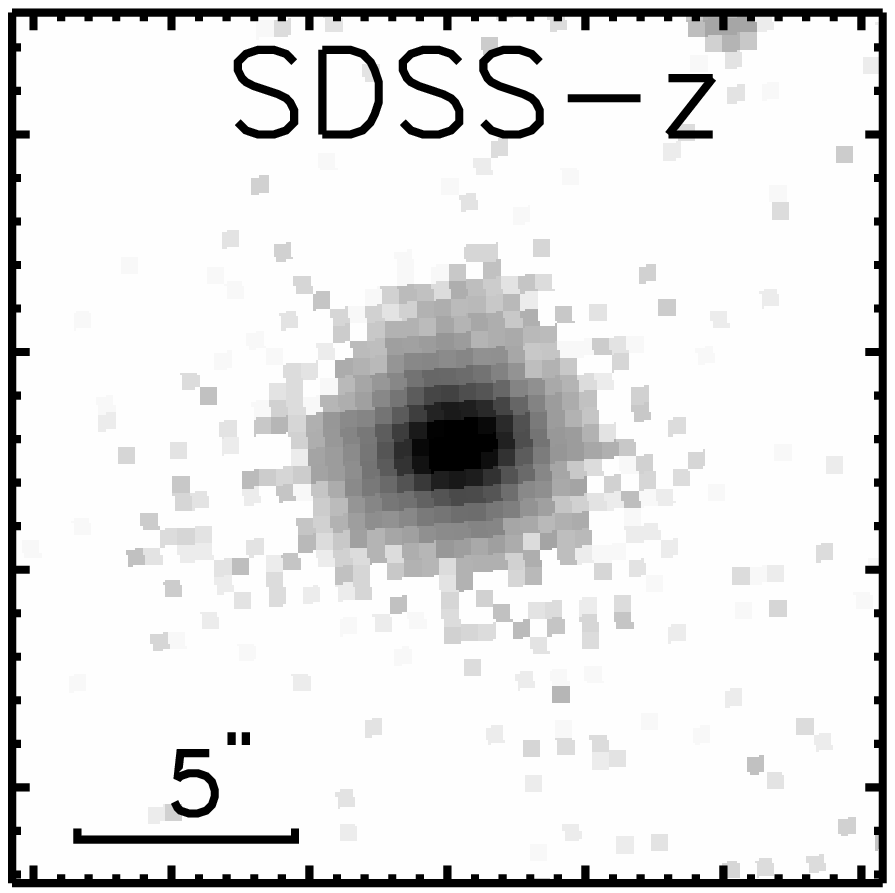}
\hspace{-8pt}
\includegraphics[width=0.33\linewidth]{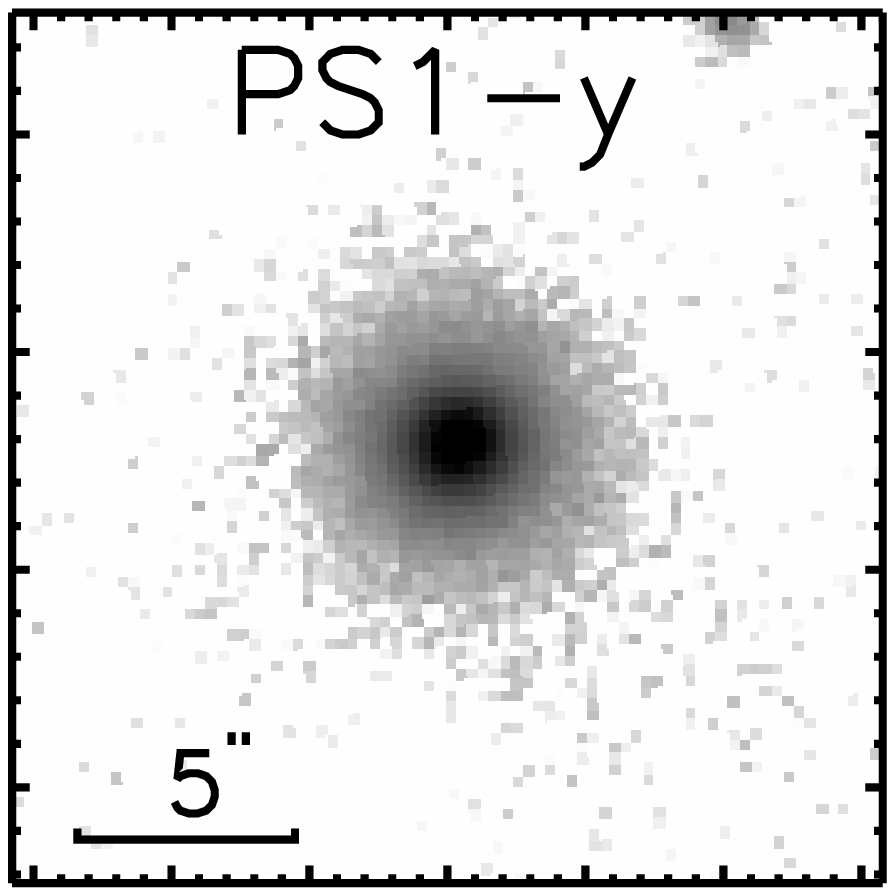}
\end{minipage}
\begin{minipage}[b]{0.305\textwidth}
\centering
\includegraphics[width=0.94\linewidth]{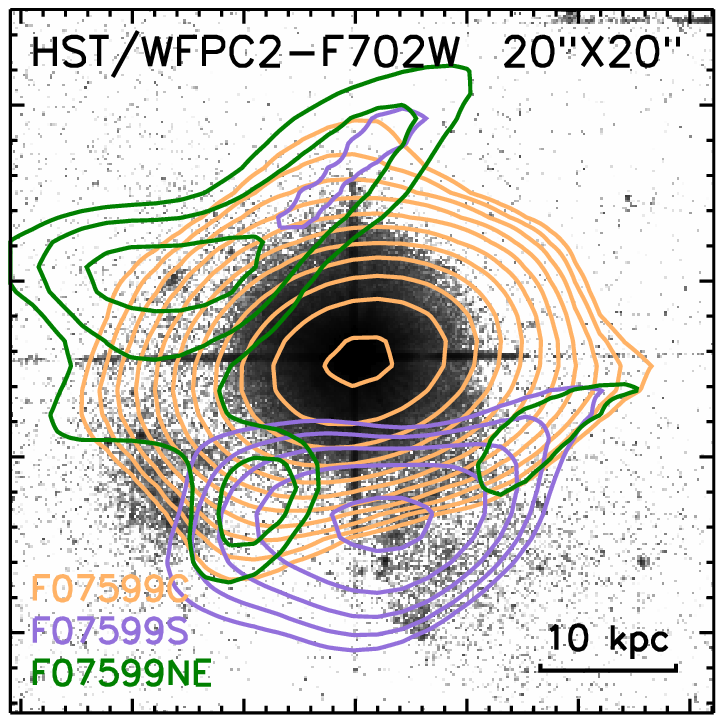}
\end{minipage}
\caption{Left: optical images of IRAS F07599+6508. From left to right and top to bottom, we show the SDSS $u$-, $g$-, $r$-, $i$-, $z$-band, and Pan-STARRS1 $y$-band images. Right: overlay of CO(1$-$0) contours (colored) of the three components on the HST/WFPC2 F702W band image. The CO(1$-$0) contours are at the same levels as in Figure~\ref{fig:3comps}. The optical diffraction spikes in the middle of the object reflect the central luminous point source of AGN that is significantly brighter than its host galaxy. Both the SDSS $gri$ and HST optical images show clear signs of extended structure to the southwest and southeast with respect to the main source.  \label{fig:optical} }
\end{figure*}
%%%%%%%%%%%%%%%%%%%%%%%%%%%%%%%%%%%%%%%%

Figure~\ref{fig:optical} shows the multi-wavelength images of IRAS F07599+6508 in SDSS $u$-, $g$-, $r$-, $i$-, $z$-, and PanSTARRS1 $y$-bands, and an overlay of the CO(1$-$0) emission of the three decomposed components on the HST/WFPC2 F702W optical image. The optical morphologies shown in $gri$ and HST images are found to be complex with tidal-like features, indicative of an asymmetric stellar distribution, while the ultraviolet (UV; $u$-band at $\sim$ 0.35 $\mu$m) and near-IR ($z$- and $y$-bands at $\sim$ 1 $\mu$m) emission show more regular and compact distribution. A consistent morphology structure in the UV band has been presented by \citet{surace00}. The tidal features seen in the optical is probably connected with an ongoing or past merging event in IRAS F07599+6508. Moreover, the absence of tidal tails observed in the near-IR bands is likely to suggest that the emission from optical tidal tails may trace the massive, young stellar population formed at recent epochs, since the near-IR emission is mostly sensitive to the longer-lived, low-mass stars. If this is indeed the case, it will also explain the morphology observed in UV band where the emission is largely attenuated in dust obscured regions. 

Interestingly, by comparing the distribution of CO molecular gas and stellar emission (see the right panel of Figure~\ref{fig:optical}), we find that the tidal feature in the optical is coincident with the CO substructure F07599S. This provides evidence that the CO feature observed in the S region may be a real structure. Moreover, the CO emission also show an extension to the southeast (SE), the same direction as the other tidal-tail-like feature observed in the optical images, although the CO extended structure toward the SE appear in only a few adjacent channels (e.g., channel maps at the velocity of about $-$178 \kms\ and $-$58 \kms ; see Appendix~\ref{appdix:channel}). Meanwhile, a faint peak coincident with the location of the CO peak in F07599NE is identified in the HST image, indicative of a possible association.

\subsubsection{CO kinematics}\label{subsect:kinematics}

\begin{figure*}[htbp]
\centering
\includegraphics[width=0.75\linewidth]{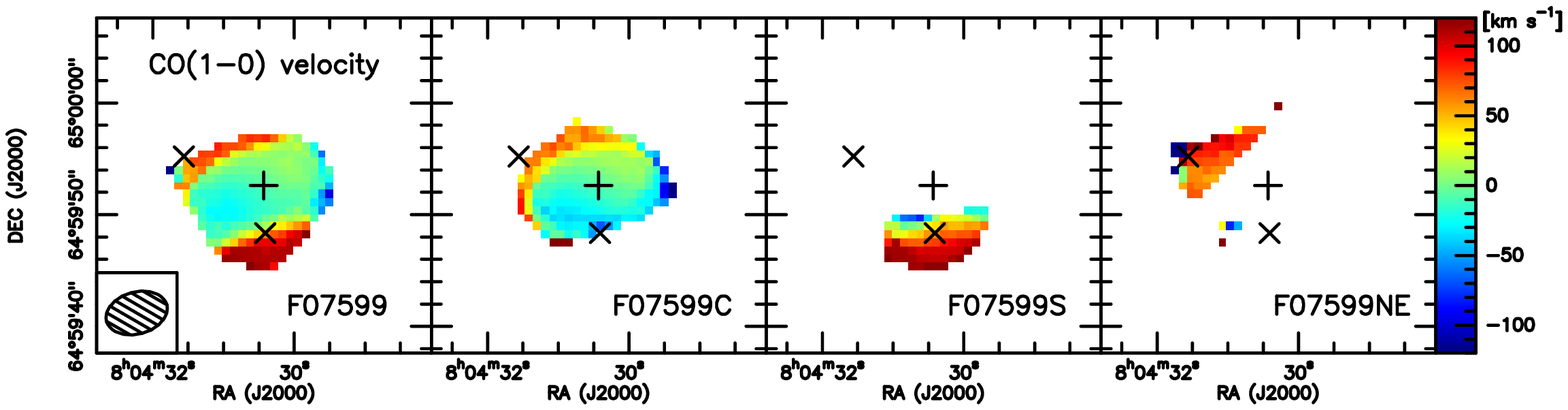}
\includegraphics[width=0.75\linewidth]{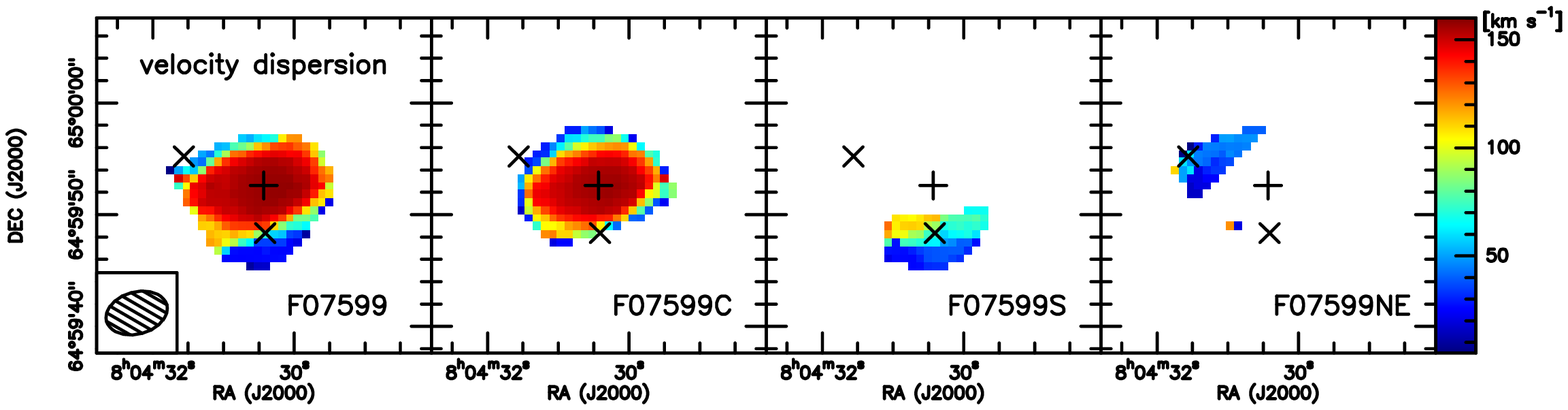}
\caption{CO(1$-$0) velocity field (top) and velocity dispersion (bottom) maps for the whole galaxy IRAS~F07599+6508 (first column) and the three subcomponents, F07599C (second column), F07599S (third column), and F07599NE (fourth column). The CO velocity and velocity dispersion maps are calculated using pixels deteccted at $\geqslant 3 \sigma$ and are masked where the integrated intensity is $<3\sigma$ in each component. The plus sign marks the peak of 3~mm continuum emission, while the two crosses indicate the CO peak of F07599S and F07599NE, respectively. The synthesized beam is shown in the bottom-left corner of the first column panel. \label{fig:mom}}
\end{figure*}

In Figure~\ref{fig:mom} we show the velocity and velocity dispersion maps of CO(1$-$0) emission within [$-$330,+330] \kms\ velocity range for the whole galaxy IRAS~F07599+6508 and the three decomposed components, respectively. These maps were created by applying a flux threshold of 3$\sigma$. The velocity map of  the whole galaxy IRAS~F07599+6508 shows irregular structure, indicative of complex velocity distribution. For the main component F07599C that is decomposed from the whole system, a systematic velocity gradient seems to be present in the north-south direction, from about $-$70 to +70 \kms, although the gas at the edge to the southeast and west appear to be disturbed. This implies that the bulk of molecular gas is likely rotationally supported. However, the spatial resolution of our data is not adequate to unambiguously interpret the observed velocity gradient as due to disk rotation. An alternative possibility could be that a velocity gradient and broad line profile may originate from an ongoing merger, as the velocity gradient is observed roughly along the direction connecting F07599C and F07599S. For F07599S, a velocity gradient of $\sim$180 \kms (from about $-$70 to +110 \kms) is also observed in the north-south direction, while no well-defined structure is visible in the velocity map of F07599NE, in which the velocity is positive around 60$-$90 \kms\ with respect to the QSO redshift.

%%%%%%%%%%%%%%%- Fig-6 -%%%%%%%%%%%%%%%%
%%%%%%%%%%%%%%%%%%%%%%%%%%%%%%%%%%%%%%%%

If we assume a rotating disk geometry for the main component F07599C, we can estimate the dynamical mass within the CO-emitting region as $M_{\rm dyn}/M_\odot \approx 2.3\times 10^5 v^2_{\rm cir} R$ \citep[e.g.,][]{wang10}, where $R$ is the CO disk radius of 4.6 kpc (0.75 times the CO FWHM major axis from visibility fitting; see Table~\ref{tab:results}) and $v_{\rm cir}$ is the maximum circular velocity of the gas disk in \kms. The $v_{\rm cir}$ is estimated as $v_{\rm cir}=0.75 \Delta v_{\rm FWHM}/{\rm sin}i$ \citep{ho07}, where $i$ is the inclination angle between the polar axis of disk and the line of sight and can be roughly calculated from the CO minor and major axis ratio ($a_{\rm min}/a_{\rm maj}$; see Table~\ref{tab:results}), i.e., $i={\rm cos^{-1}}(a_{\rm min}/a_{\rm maj})=45^\circ$. The derived $M_{\rm dyn}$ is $2.1\times 10^{11}\ M_\odot$ with a $v_{\rm cir}$ of 450 \kms\ for the primary disk F07599C. We stress that this estimate of dynamical mass can be significantly affected by the large uncertainties of source size, inclination angle, and the assumption of an inclined disk geometry. The total amount of molecular gas in the host galaxy is $M_{\rm tot,mol}\sim1.1\times 10^{10}\ M_\odot$ by adopting a classic ULIRG-like CO-to-H$_2$ conversion factor of $\alpha_{\rm CO}=0.8\ M_\odot\ {\rm (K\ km\ s^{-1}\ pc^{-2})^{-1}}$. We obtain a molecular gas fraction of $\sim0.05$ for IRAS~F07599+6508 by assuming the gas fraction can be approximated as the ratio of gas to dynamical mass, slightly lower than those (median value of 0.22$\pm$0.04) of IR QSOs measured by \citet{tan19}, and the local ULIRGs as well \citep[][]{downes98}. Here we note the large uncertainties involved in the estimate of molecular gas mass, due to the assumption of conversion factor $\alpha_{\rm CO}$, which is uncertain and varies significantly from source to source \citep{bolatto13}. Recent studies show evidence for higher $\alpha_{\rm CO}$ values in (U)LIRGs that are affected by outflows if a substantial amount of molecular gas is in dense, gravitationally bound states \citep[e.g.,][and references therein]{cicone18a}.

The velocity dispersion map shows $\sigma_{\rm v} \sim$160 \kms\ at the center of IRAS~F07599+6508, which is significantly higher than the typical peak dispersion of $\sim$80 \kms\ observed in local IR QSOs \citep{tan19}, and $z\gtrsim$6 QSOs ($\sim$130 \kms) as well \citep{neeleman21}. A plausible explanation could be related to the beam smearing effect that increases the observed dispersion, since the spatial angular resolution of our NOEMA observations is about an order of magnitude lower than the ALMA observations ($\sim 0.\arcsec 5$) shown by \citet{tan19}. Assuming a uniform intrinsic dispersion across the galaxy disk of IRAS~F07599+6508, we then estimate a dispersion of $\sigma_{\rm v} \sim 70-100$ \kms, which is measured in the outer region of F07599C where the beam smearing is considered to be less severe \citep{davies11}.  The velocity dispersion map of F07599S shows an increase in the north direction with a peak dispersion of $\sim$110 \kms\ which is in the overlap region between F07599C and F07599S, while the gas in F07599NE appears less perturbed with dispersion of $\sim 30-60$ \kms. The increased velocity dispersion observed in the overlap region could be interpreted as a turbulence-dominated shock region induced by a merger or a possible CO outflow component. Higher resolution observations are needed to better constrain the intrinsic velocity dispersion.

\subsection{Radio Continuum Emission}

In star-forming galaxies, the radio emission is dominated by synchrotron radiation from relativistic electrons accelerated in supernovae remnants of massive stars. It is largely optically thin and unaffected by dust extinction, and therefore an excellent probe of very recent star formation activity in galaxies \citep{condon92}. For the radio-quite (RQ) AGNs, which represent the majority of the AGN population, the origin of radio emission could be connected with a wide range of possible mechanisms, namely, star formation, AGN-driven wind, jets, and accretion-disk coronae \citep{panessa19}.

\begin{figure}[htbp]
\centering
\includegraphics[width=0.95\linewidth]{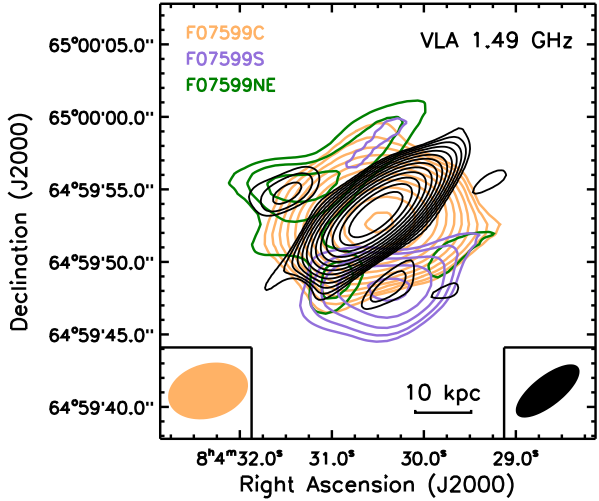}
\caption{VLA 1.49 GHz contour (black) maps of IRAS~F07599+6508 overlaid on the CO(1$-$0) contour (colored) maps of the three components. The CO(1$-$0) contours are at the same levels as in Figure~\ref{fig:3comps}. The contours of 1.49 GHz continuum emission start at 2$\sigma$ (1$\sigma$ noise level is 74.9 $\mu$Jy beam$^{-1}$) and increase by a factor of 1.5. The solid orange and black ellipses in the bottom left and right corners are the beam size of CO and 1.49 GHz maps, with angular resolutions of 5.\arcsec61$\times$3.\arcsec77 and 5.\arcsec33$\times$2.\arcsec08, respectively. \label{fig:rc-lband}}
\end{figure}

Figure~\ref{fig:rc-lband} shows the VLA 1.49 GHz contour maps overlaid on the NOEMA CO(1-0) data that is decomposed into three components. Along the direction of the minor axis of synthesized beam of 1.49 GHz observations, two clumpy structures outside the central bright compact radio source are potentially detected at $\sim 3\sigma$ and $\sim 4\sigma$ level to the S and NE direction, respectively. The radio peaks of both the central compact source and the two clumps are well aligned with the CO peaks of the three decomposed components, reinforcing the fact that the faint CO features observed in the S and NE direction are very likely to be real.

%%%%%%%%%%%%%%%- Fig-7 -%%%%%%%%%%%%%%%%
%%%%%%%%%%%%%%%%%%%%%%%%%%%%%%%%%%%%%%%%

A 2D Gaussian fit to the 1.49 GHz image returns a peak flux density of 32.33$\pm$0.10 mJy beam$^{-1}$ and a consistent integrated flux density of 32.43$\pm$0.19 mJy, indicative of a relatively compact structure of radio emission. This measurement agrees well with the 1.4 GHz flux density of 39.5~mJy derived from the NRAO VLA Sky Survey \citep[][]{condon98}. IRAS~F07599+6508 is additionally detected at 4.86, 8.44, 14.94, and 22.46 GHz by VLA. The maps and the flux measurements at these radio frequencies are shown in Appendix~\ref{appdix:radio}. At the redshift $z=0.1486$, observations at the radio frequencies listed in Table~\ref{tab:radio} sample $\sim$1.7, 5.6, 9.7, 17.2, and 25.8 GHz in the rest-frame, respectively. In the following section, we adopt the rest-frame frequencies for clarity.

\begin{figure}[t!]
\centering
\includegraphics[width=0.85\linewidth]{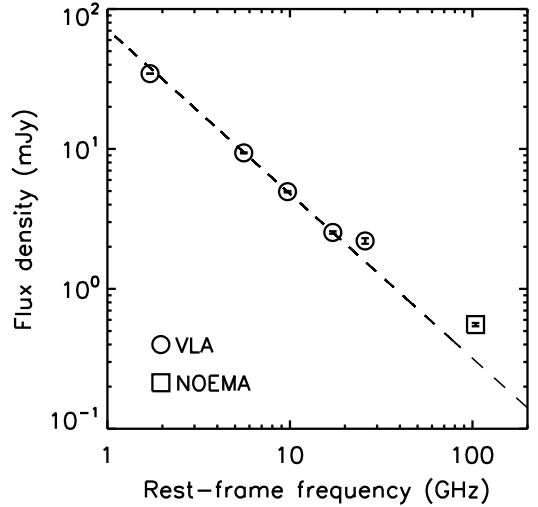}
\caption{The radio-to-mm spectral energy distribution of IRAS~F07599+6508 based on the data from VLA (circle) archive and NOEMA (square). \label{fig:rc-sed}}
\end{figure}

Figure~\ref{fig:rc-sed} shows the SED of the continuum emission at radio and mm bands in IRAS~F07599+6508. At GHz frequencies, the flux density $S_\nu$ in terms of the observed frequency $\nu$ is generally defined as $S_\nu \propto \nu^\alpha$, with $\alpha$ the radio spectral index. The spectral indices measured from 1.7$-$5.6, 5.6$-$9.7, 9.7$-$17.2, and 17.2$-$25.8 GHz are $\alpha_{5.6}^{1.7}=-1.10\pm0.09$, $\alpha_{9.7}^{5.6}=-1.16\pm0.22$, $\alpha_{17.2}^{9.7}=-1.18\pm0.26$, and $\alpha_{25.8}^{17.2}=-0.34\pm0.78$, respectively. It is clear that the spectrum turns over from steep to flat at 17.2$-$25.8 GHz, which is consistent with the expectation that the contribution of thermal emission increases at higher frequencies. At low frequencies, the spectral index $\alpha_{5.6}^{1.7}$ is slightly flatter than those at higher frequencies, $\alpha_{9.7}^{5.6}$ and $\alpha_{17.2}^{9.7}$, both of which show almost identical values. The flatter index observed at lower frequencies has been interpreted as a result of optical depth effects that the radio spectrum typically shows a sharp decline due to free-free absorption \citep[e.g.,][]{condon92,clemens10}. In contrast, the flux at larger frequency should be much less affected \citep[e.g.,][]{clemens08}. Therefore, we assume a synchrotron (non-thermal) spectral index of $\alpha^{\rm NT}=-1.18$, equivalent to the steepest part of the radio spectrum measured between 9.7 and 17.2 GHz (see the dashed line in Figure~\ref{fig:rc-sed}). This slope is slightly steeper than a normal synchrotron spectral index of $\alpha=-0.8$ derived for a normal star-forming galaxy \citep{condon92}, but consistent with the slopes measured for RQ AGN \citep[e.g.,][]{behar15}.

We extrapolate the contribution of non-thermal emission from $\alpha_{17.2}^{9.7}$ at the NOEMA frequency of 90.5 GHz ($\sim$ 103.9 GHz in the rest-frame) and subtract from the observed flux density. In addition, subtracting off an estimate of the thermal dust contribution to the 103.9 GHz, which is derived from a grey-body model fit to the $Hershel$/SPIRE data with an emissivity spectral index $\beta$=1.6 \citep{clements18}, we obtain an estimate of free-free emission component of 0.17 mJy. This implies a free-free radio fraction at 103.9 GHz of $\sim$30\%. A similar procedure applied to 25.8 GHz data leads to a free-free fraction of $\sim$30\% estimated at this frequency. This is consistent with the median thermal fraction of 20-30\% at 22 GHz measured for local ULIRGs \citep{vega08}.

We derive an extrapolated 1.4 GHz flux density of 43.4 mJy from the observed rest-frame 1.7 GHz flux density by assuming a radio power-law index of -1.18. Using the flux densities measured at IRAS 60 and 100 $\mu$m, we calculate a FIR-radio flux ratio\footnote{The FIR-radio flux ratio $q$ is defined as the ratio of the FIR luminosity (40-120 $\mu$m) to the radio power \citep{condon92}:\\ $q \equiv$ log$_{10}\left(\frac{L_{\rm FIR}}{3.75\times10^{12}\ \rm{W}}\right)$-log$_{10}\left(\frac{L_{\rm 1.4 GHz}}{\rm{W\ Hz^{-1}}}\right)$} $q=1.67$. This implies a linear ratio about 4 times smaller than the average value of $q=2.34\pm 0.01$ found for local star-forming galaxies \citep{yun01}, possibly suggestive of an additional contribution from compact AGN source that boosting the radio flux.

%%%%%%%%%%%%%%%- Fig-8 -%%%%%%%%%%%%%%%%
%%%%%%%%%%%%%%%%%%%%%%%%%%%%%%%%%%%%%%%%

\section{Discussion}\label{sec:diss}

The main results of our NOEMA observations are the detection of 3~mm continuum emission and the multiple spatially decomposed CO components in the host galaxy of IRAS~F07599+6508. By combining with multi-wavelength data, we begin the discussion by looking into the properties and the possible origin of these components.

\subsection{Nature of the 3mm Emission}

The 3~mm continuum emission of IRAS~F07599+6508 is resolved with a deconvolved size of 1.\arcsec5 ($\sim 4.0$ kpc). Assuming a non-thermal spectral index of $\alpha^{\rm NT}=-1.18$, we find that about 30\% of the emission at 3~mm arises from a thermal free-free component. This is significantly lower than the thermal fraction of $\sim70\%$ at 3~mm expected for normal star-forming galaxies \citep{condon92}, indicating that the nonthermal synchrotron component from AGN may contribute substantially to the 3~mm emission in IRAS~F07599+6508. This could be expected since it does not follow the FIR-radio correlation characteristic of star-forming galaxies, with a ratio of $L_{\rm FIR}/L_{\rm 1.4 GHz}$ being about four times lower. A similar situation has been observed in Mrk~231, the nearest IR QSO with a compact radio core\citep[e.g.,][]{lonsdale03}, where a rather low  thermal fraction is observed at 33 GHz \citep{barcos-munoz17}. As mentioned above, an additional nonthermal component from AGN could contribute to the emission at high radio frequencies. An alternative explanation for the observed low thermal fraction is connected with the possibility that, the production of free-free emission might be suppressed due to the absorption of ionizing stellar photons by dust that concentrates in the dense starburst regions.

Taking the 2-10 keV X-ray luminosity from \citet{brightman11}, we find a radio ($\sim3$ mm) luminosity to X-ray luminosity ratio of $L_{\rm mm}/L_{\rm X}\sim 3\times10^{-2}$ for IRAS~F07599+6508. This value is about two orders of magnitude lower than those of radio-loud (RL) AGNs, which typically have powerful relativistic jets that shine in the radio band, but close to the ratios ($10^{-2}-10^{-4}$) measured for RQ AGNs \citep{behar15,panessa19}, suggestive of a different origin of radio emission compared with RL QSOs. We note that the ratio of $L_{\rm mm}/L_{\rm X}\sim 3\times10^{-2}$ derived here could be an overestimated value, given that the AGN embedded in the galaxy IRAS~F07599+6508 is known to be heavily obscured and thus a portion of X-ray photons may be absorbed \citep[see][]{luo14,la-caria19}. A connection between the radio emission and the nuclear accretion disk has been proposed to explain the correlation of the radio luminosity $L_{\rm mm}$ and the X-ray luminosity $L_{\rm X}$ observed in RQ AGNs \citep[e.g.,][]{laor08}. More evidences for an X-ray corona (i.e., hot plasma with a temperature of $\sim10^9$ K) origin of radio emission in RQ AGNs have been presented by the detections of millimeter excess \citep[e.g.,][]{behar15,inoue18}. These are in line with the theoretical prediction for a radio synchrotron component from X-ray corona in the vicinity of black hole accretion disks contributing to the emission in the millimeter band \citep[see][]{inoue14,raginski16}. Nevertheless, a conclusive evidence for the coronal radio emission would be the detection of the radio/millimeter and X-ray variability correlation similar to that observed in stellar coronae \citep{panessa19}.

The thermal free-free emission arises directly from H{\sc ii} regions ionized by massive stars and its intensity is proportional to the ionizing photon rate from young star-forming regions, and thereby provide an excellent diagnostic for the current SFR of galaxies \citep{condon92}. Taking the free-free emission flux measured at rest-frame 103.9 GHz and the free-free SFR calibration given in \citet{murphy12}, we calculate an SFR of 77 $M_\odot$ yr$^{-1}$ for IRAS~F07599+6508 by assuming a thermal spectral index of $-$0.1 and an electron temperature of $T_{\rm e}=10^4$~K. Compared to the IR-derived SFR of 280 $M_\odot$ yr$^{-1}$ that is estimated based on the heated dust component \citep{hao08}, a lower SFR measured by the free-free emission could be expected since free-free emission is sensitive to massive stars with ages $\lesssim$10 Myr and consequently probe the very recent star formation activity. In addition, the free-free-derived SFR would be underestimated if a significant fraction of ionizing photons are absorbed by dust. Alternatively, it could be that a portion of IR luminosity arise from the dust heating by nuclear luminous AGN, which would lead to an overestimate of SFR. Indeed, we obtain an SFR of 150 $M_\odot$ yr$^{-1}$ using the calibration given by \citet{murphy12} by correcting for the AGN contribution to the bolometric luminosity \citep[$\alpha_{\rm AGN}=0.75$;][]{veilleux09}, which is close to the free-free SFR estimate. All of these including the uncertainties (e.g., variations in the actual $T_{\rm e}$) on the estimation of free-free-derived SFR may explain the discrepancy in SFR estimates.

\subsection{Origin of the off-center CO components}

By exploiting the CO line cube of IRAS~F07599+6508, we have identified two off-center CO subcomponents, F07599S and F07599NE, at a significance level of 16.0$\sigma$ and 6.4$\sigma$ in the intensity map, respectively. Figure~\ref{fig:3comps} shows the CO(1$-$0) spectra and maps extracted for these two subcomponents and the main component F07599C as well. F07599S and F07599NE are located to the south and northeast of the disk component with a projected distance of $\sim$11.4 and 19.1 kpc and with a fraction of 9\% and 4\% of CO flux contributed to the total CO flux, respectively. These two CO subcomponents are not detected in 3mm continuum emission. For the main component F07599C, the systematic CO velocity gradient observed may be a signature of a rotating disk structure. If confirmed, the source size of $\sim 6.1$~kpc measured would suggest that the CO disk is very extended, which is in sharp contrast with local ULIRGs where the majority of molecular gas is centrally concentrated within a radius of $\sim$1~kpc \citep{downes98,wilson08}.

The optical images of IRAS~F07599+6508 show evidence for a tidal-tail-like structure (see Figure~\ref{fig:optical}), suggestive of a recent merging event in IRAS~F07599+6508. It is interesting to note that the tidal feature shown in the optical image is spatially coincident with the CO feature detected in the S region. In addition, a faint peak coincident with the location of the CO peak in F07599NE is identified in the HST optical image, likely associated with the origin of the CO plume feature observed in the NE direction. It can be seen from the channel maps in Appendix~\ref{appdix:channel} that the CO emission is extended to the S and NE direction, suggesting that the CO emission identified in the extranuclear region of F07599S and F07599NE are very likely to be real components. Furthermore, the two potentially detected clumpy structures revealed in the 1.49 GHz radio image are found to align well with the peaks of F07599S and F07599NE (see Figure~\ref{fig:rc-lband}), as well as the faint substructures detected at 4.86 and 8.44 GHz (see Appendix~\ref{appdix:radio}), supporting for the reality of such CO feature. 

In Figure~\ref{fig:3comps} we have shown that the CO(1$-$0) spectrum extracted for F07599S is best-fitted with a two-component (narrow and broad) Gaussian. The detection of narrow (FWHM = 100$\pm$10 \kms) feature may indicate the presence of extended dynamical ``quiescent'' CO gas. For the broad component, the Gaussian fit gives an FWHM of 500$\pm$80 \kms, slightly larger than the line FWHM (430$\pm$10 \kms) of the primary component F07599C. Together with the shift in velocity ($\sim$100 \kms) and the tidal-tail-like structure observed in the optical image, it is likely that F07599S represents a separate component from main galaxy, and has a merger origin. One of the possibilities is that F07599S may trace the emission from a small obscured companion to the host galaxy of IRAS~F07599+6508. The non-detection of UV emission in F07599S show evidence for the large attenuation by dust (see Figure~\ref{fig:optical}). More evidence for a merger origin is given by the increased velocity dispersion observed in the overlap region between F07599S and the main component, which could be interpreted as a turbulence-dominated shock region induced by an interaction. 

For F07599NE, the best-fit of the CO(1$-$0) line is a single Gaussian with both FWHM and peak velocity consistent within the errors with the main component. The significant lower velocity dispersion observed in this component imply that F07599NE may have a different origin from F07599S. Similar gas substructure has been observed in high-$z$ starbursts in [C {\sc ii}] emission but with a much smaller physical separation \citep[$\sim 2$ kpc;][]{tadaki20}, where the formation of subcomponent is expected to be associated with a gravitationally unstable disk or shocks. By contrast, the large physical separation (projected distance of $\sim 19$ kpc) observed between F07599NE and the galaxy nucleus indicates that the CO plume feature observed in the NE direction could be a separate component, likely associated with the tidal debris reminiscent of a previous merger. Similar results have been seen in the IR QSO IRAS~F23411+0228, where an off-center clumpy structure is detected in both CO line and 3~mm continuum maps and is found to align  with a faint peak identified in the HST image, which was interpreted as either an ongoing merger or a clumpy star-forming region at large radii \citep{tan19}. \citet{cicone15} have shown a very extended (up to $r\sim$ 30 kpc) [C {\sc ii}] cold gas component observed in a $z>6$ QSO host galaxy and they interpret that the extended [C {\sc ii}] emission may arise from a different ISM phase than dense photon-dominated regions. 

We note that there is a source accompanying IRAS~F07599+6508 in multi-wavelength data set (from GALEX near-UV to WISE 3.4 $\mu$m bands; see Appendix~\ref{appdix:multidata}), with a distance of 17.\arcsec5 (projected distance of $\sim$ 45 kpc) to the northwest. This source has a cross-identification as WISEA J080428.28+650001.7 (hereafter J0804+6500) in the NASA/IPAC\footnote{\url{http://ned.ipac.caltech.edu}} Extragalactic Database. There is no optical spectroscopic data available for this object. In Appendix~\ref{appdix:multidata}, we show the NOEMA spectrum extracted from the position of J0804+6500 by fitting the visibilites with a point source. We do not detect any significant emission in either continuum or line from this source. With the data available, it is not possible to assess whether J0804+6500 is physically related with the IRAS~F07599+6508 system, or is only a projected companion source. The detection of line emission from this source would help us measure the redshift and further determine whether it is an interacting companion of IRAS~F07599+6508.

Together with the CO clumpy structures of local IR QSOs revealed with ALMA high resolution observations \citep{tan19}, the multiple components revealed in the CO maps show evidence for the complex morphology that may associate with the intense star formation and AGN activities. However, additional deep CO observations with higher angular resolution are needed for confirming the faint features observed in F07599S and F07599NE and testing the corresponding hypothesis.

%%%%%%%%%%%%%%%- Fig-9 -%%%%%%%%%%%%%%%%
\begin{figure*}[htbp]
\centering
\includegraphics[width=0.4\linewidth]{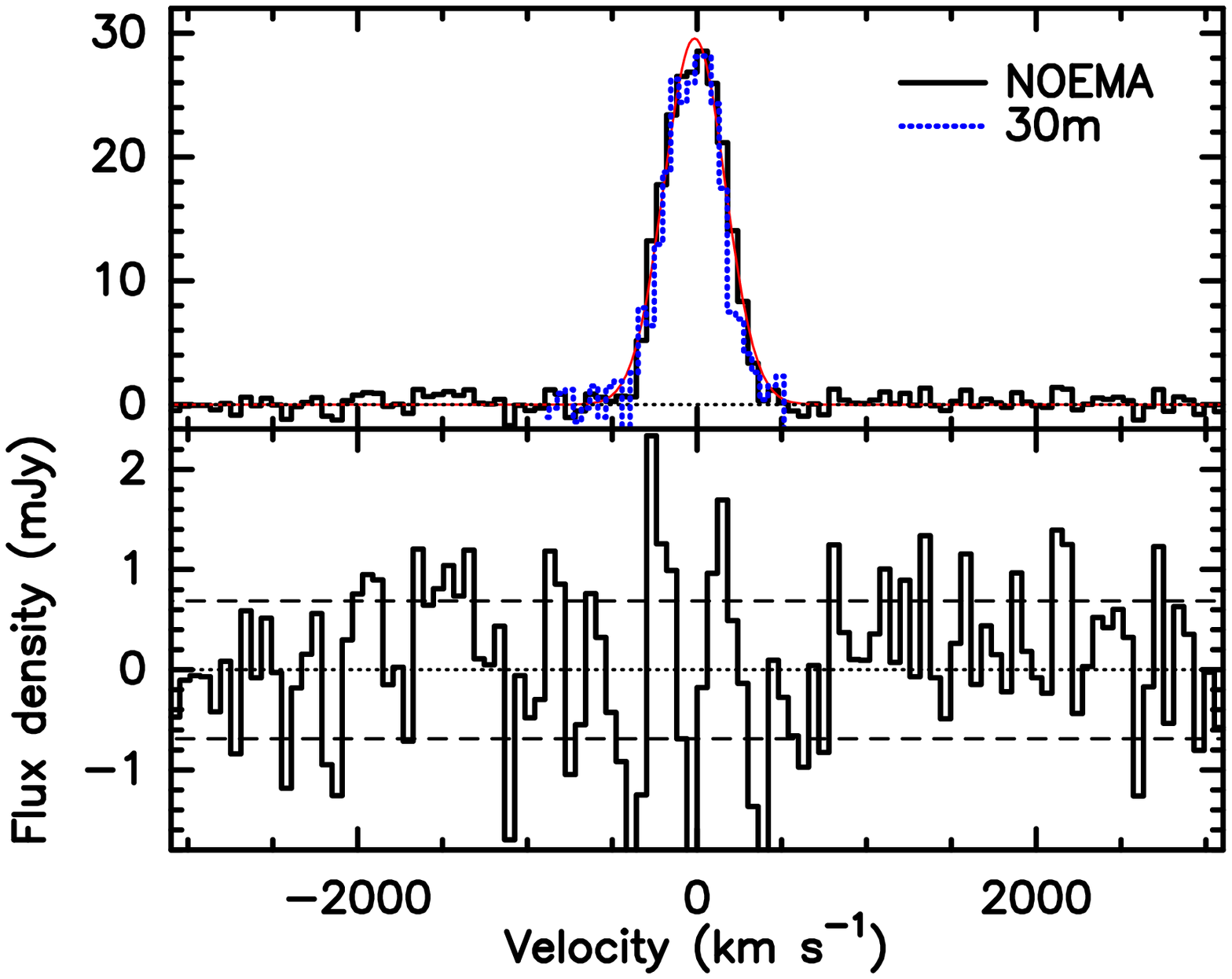}
\hspace{15pt}
\includegraphics[width=0.3\linewidth]{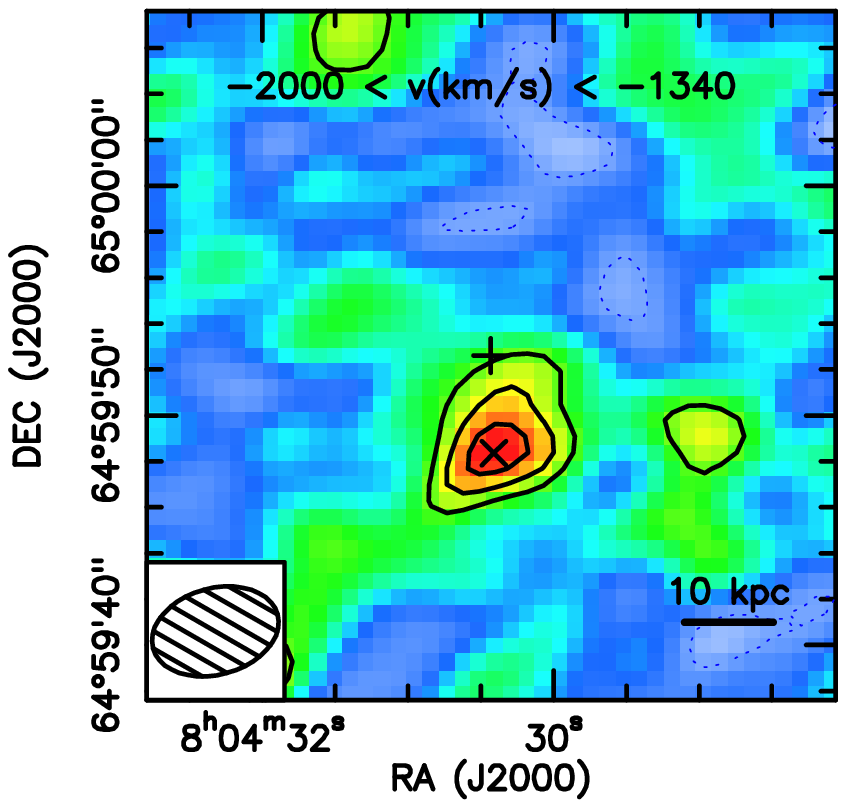}
\caption{Left: continuum-subtracted spectra of the CO(1$-$0) emission line (top) and the residuals (bottom) after removing a Gaussian fit (red line in the top panel) to the CO spectrum of IRAS~F07599+6508. In the top panel, the NOEMA CO spectrum (black) is obtained by summing up the flux of the three decomposed components in each channel (see the left panel in Figure~\ref{fig:3comps}), while the IRAM 30m CO spectrum (blue) is also plotted for a comparison. In the bottom panel, the dashed lines denote the 1$\sigma$ level. The spectra are binned to a velocity resolution of $\sim$60~\kms. Right: contours from CO(1$-$0) dirty maps integrated over the velocity range of [$-$1340,$-$2000] \kms. Contours starts at $\pm 2 \sigma$ and increase in steps of 1$\sigma$ (81~$\mu$Jy~beam$^{-1}$) with positive (negative) contours shown as solid (dotted) lines. The plus sign and cross indicate the CO peak positions of F07599C and F07599S, respectively. The synthesized beam is shown in the bottom-left corner. \label{fig:high-velo}}
\end{figure*}
%%%%%%%%%%%%%%%%%%%%%%%%%%%%%%%%%%%%%%%%

\subsection{Comparison of molecular gas with Mrk~231}\label{mrk231}

As noted previously in Section~\ref{sec:intro}, IRAS~F07599+6508 and Mrk~231 share many common or similar properties in multi-wavelength, i.e., X-ray, UV, optical, and IR, and have been proposed as local templates of candidate ULIRG-to-optical QSO transition objects. High-resolution CO observations of Mrk~231 show that the molecular gas is concentrated in the central $\sim$1 kpc rotating disk \citep{downes98,cicone12,feruglio15}, though O$_2$ emission might be present over $\sim$10 kpc \citep{wang20}. Compared with Mrk~231, the  molecular gas of IRAS~F07599+6508 is also likely distributed in a disk with systematic velocity gradient in the main component, but much more extended with a size of $\sim$6.1 kpc. In addition, the CO line width of IRAS~F07599+6508 is $\sim$430 \kms\ FWHM, much broader than that of 190 \kms\  measured for the core component of Mrk~231 \citep{downes98}. The CO(2$-$1) to CO(1$-$0) line luminosity ratios derived for IRAS~F07599+6508 and Mrk~231 are both $\sim 0.8$ \citep{xia12,cicone12}, indicative of modest sub-thermal molecular gas in their hosts.

The detection of massive, highly energetic, and kpc-scale molecular outflows in Mrk~231 show direct observational evidence for AGN feedback onto the host galaxy \citep[e.g.,][]{feruglio10,feruglio15,aalto12,cicone12}, consistent with the AGN-galaxy co-evolutionary models which propose that the outflow feedback is an important process in quenching the star formation activity in massive galaxies \citep[e.g.,][]{DiMatteo05,hopkins06}. The CO(1$-$0) wing-to-peak ratio in Mrk~231 is $\sim$2.5\%, in agreement with the ratio of 1-8\% typically measured for local ULIRGs \citep{feruglio10,cicone12,cicone14}.

The presence of molecular outflow in IRAS~F07599+6508 has been confirmed by the detection of OH 119 $\mu$m P-Cygni profile with velocities $<$ 1000 \kms\ \citep{veilleux13}. A fast wind has also been seen in this system in the ionized gas as traced by the broad blueshift optical spectra (H$\alpha$ and H$\beta$) and in the neutral atomic phase with broad absorption observed in Na~{\sc i}~D line \citep[e.g.,][]{zheng02,rupke17}.

In the left-top panel of Figure~\ref{fig:high-velo} we show the continuum-subtracted NOEMA CO(1-0) spectrum of the whole galaxy of IRAS~F07599+6508, which is obtained by summing up the flux of the three decomposed components in each channel, compared with the CO(1-0) line observed with the IRAM 30m \citep{xia12}. The CO peak flux measured by NOEMA is found to agree well with the IRAM 30m data, suggesting that little emission is filtered out by our interferometric observations. The residual spectrum after removing a Gaussian fit to the core component of the CO spectrum is shown in the left-bottom panel of Figure~\ref{fig:high-velo}. A noise level (1$\sigma$) of 0.68 mJy in each $\sim$ 60 \kms\ channel was measured, corresponding to 2.3\% of the peak flux density. Scaling to the signal strengths from Mrk~231 would lead to an estimated flux density of the CO broad emission of $\sim$ 0.75 mJy for IRAS~F07599+6508, comparable to the noise level of our NOEMA data.

It is interesting to note that several consecutive channels in the blue-shifted side of the CO spectrum (see Figure~\ref{fig:high-velo}, left-bottom) show flux density higher than 1$\sigma$. We averaged the visibilities of the high-velocity component in the velocity range [$-$1340, $-$2000] \kms\ and fitted a point-source model. The peak of high-velocity CO emission is located between the F07599C and F07599S with a projected distance of 4.\arcsec3 from the galaxy nucleus (see Figure~\ref{fig:high-velo}, right). The flux derived from \texttt{uv\_fit} is 0.228$\pm$0.054 Jy \kms, corresponding to a detection significance of 4.2$\sigma$. However, this does not satisfy the criterion defined by \citet{cicone14}, i.e., CO emission with velocities $>500$ \kms\ is detected with a significance of at least 5$\sigma$, for claiming the detection of a molecular outflow. The OH outflow counterpart in CO emission is not significantly detected in our NOEMA data, although there is a hint of some flux at very high blueshifted velocities in the overlap region between F07599C and F07599S (see Figure~\ref{fig:high-velo}), where the extracted CO spectrum shows a clear decomposition into broad plus narrow components (see the CO spectrum of F07599S in Figure~\ref{fig:3comps}). Nevertheless, it is difficult to rule out the existence of a weak CO outflow with the current data. 

Given the much extended CO disk providing fuel for active large-scale star formation in IRAS~F07599+6508, perhaps a much weaker outflow feedback in quenching star formation is expected here as compared to that of Mrk~231, where bursts of star formation only remain in the central compact core. If that is true, IRAS~F07599+6508 should be in an earlier phase than that of Mrk~231 in the sense of ULIRGs evolution in transition into the classical QSOs. We stress that additional NOEMA CO observations are required to investigate this source in detail, particularly the presence of a possible CO outflow.

\section{Summary}\label{sec:sum}

In this work, we present the deep NOEMA interferometric observations of CO(1$-$0) line and 3 mm continuum emission in the ultraluminous IR QSO IRAS~F07599+6508 at $z=0.1486$. This system shares many physical properties with Mrk~231, the nearest and most luminous IR QSO, in multi-wavelength. Both the CO line and 3mm continuum emission are resolved in our observations, and the CO gas is found to be spatially decomposed into three components. Combining these observations with results from multi-wavelength data (i.e., radio, optical, and infrared), we study the properties of the 3~mm continuum emission and the CO components spatially decomposed, and the main results are summarized as follows.

1. The 3~mm continuum emission is resolved with a deconvolved size of 1.\arcsec 5 ($\sim4.0$ kpc). Assuming a non-thermal spectral index of $\alpha^{\rm NT}=-1.18$, equal to the slope measured between 9.7 and 17.2 GHz representing the steepest part of the radio spectrum, we find that $\sim30$\% of the emission at 3~mm arises from a free-free component, significantly lower than the fraction ($\sim70\%$) expected for a normal star-forming galaxy. This suggests that the nonthermal synchrotron component from AGN may contribute substantially to the 3~mm emission. The relatively low FIR-to-radio flux ratio of $q=1.67$ derived also show evidence for an additional contribution from the central AGN that boosting the radio flux. We estimate an SFR of 77 $M_\odot\ {\rm yr^{-1}}$ based on the thermal free-free emission component at 3~mm, which is close to the IR estimate of $150\ M_\odot\ {\rm yr^{-1}}$ corrected for the AGN contribution to the bolometric luminosity.
 
2. Two off-center CO subcomponents, F07599S and F07599NE, located to the south and northeast of the main component F07599C with a projected distance of $\sim$11.4 and 19.1 kpc, respectively, were identified by model fits to the $uv$-data. More evidence for the reality of F07599S and F07599NE are given by the spatial coincidence of the tidal-tail-like feature and the faint peak in the optical images, and the potentially detected clumpy structures in the 1.49 GHz radio image with these two components. The fraction of F07599S and F07599NE contributed to the total CO flux are 9\% and 4\%, respectively. Analysis of the CO kinematics together with the multi-wavelengths data for the decomposed components suggests that the gas in the dominant component F07599C is likely rotationally supported with a systematic velocity gradient observed, while F07599S may represent a separate component associated with a merger and F07599NE probably have a different origin.

3. Careful analysis of our current CO(1$-$0) data indicates that, the significance of the high-velocity emission feature seen in the blue-shifted side of the spectrum is not sufficient for claiming the detection of a CO outflow in IRAS~F07599+6508. It should be noted that, however, the spatially broad CO core distribution with a size larger than 6 kpc is in marked contrast with the compact $\sim$ kpc CO concentration in Mrk~231, where strong molecular outflows were observed over wide velocity ranges.

Additional deeper observations with higher angular resolution are required to better constrain the CO morphology and kinematics on kpc and sub-kpc scales, and confirm the presence of high-velocity gas and further determine whether the features seen here are potential signs of outflowing gas that is escaping from the host galaxy.

\acknowledgments

We thank the anonymous referee for the constructive comments that helped improve the paper. We acknowledge the support from IRAM staff during the observations and data reduction. This work is based on observations carried out under project number S18CA with the IRAM NOEMA Interferometer. IRAM is supported by INSU/CNRS (France), MPG (Germany) and IGN (Spain).

This work was supported by National Key Basic Research and Development (R\&D) Program of China (Grant No. 2017YFA0402704), NSFC Grant No. 11803090, 11861131007, 12033004, and 11420101002, and Chinese Academy of Sciences Key Research Program of Frontier Sciences (Grant No. QYZDJ-SSW-SLH008). X.Y.X and C.N.H. acknowledge support from the NSFC (Grant No. 11733002). The work by K.K is supported by the JSPS KAKENHI Grant Number JP17H06130 and the NAOJ ALMA Scientific Research Grant Number 2017-06B.

%%% project number XYYZZ [XXX-YY]


\vspace{5mm}
\facilities{IRAM:NOEMA, VLA, Sloan, HST, PS1}


\software{GILDAS \citep{guilloteau00},  
          CASA \citep{mcmullin07},        
           AIPS \citep{vanmoorsel96}        }



\appendix

\section{Decomposition of CO emission}\label{appdix:decomp}

To illustrate the spatial decomposition of CO emission in IRAS~F07599+6508, we here describe the process of model fitting procedure. We fit the model to the $uv$-data using combinations of functions by allowing all parameters to be free. Combinations of models in a successive progression of complexity started with a point source and then circular Gaussian, and elliptical Gaussian. The simultaneous fit was rerun each time as necessary to achieve a uniform residual (i.e., no significant peak or negative regions $\gtrsim 3\sigma$) after subtraction of the models. By verifying the evolution of the uncertainties on the fitted parameters, we found that the best fit was obtained with a combination of an elliptical Gaussian (F07599C) and two point sources (F07599S and F07599NE) functions with minimum uncertainties on the fitted parameters. The residuals after subtracting an elliptical Gaussian plus two point source components are shown in the right panel of Figure~\ref{fig:residuals}. It should be noted that there are a few peaks with SNR$\sim3$ (e.g., the peaks to the northwest of F07599S and F07599NE, respectively) in the residual map, indicative of the likely presence of extended structure. These weak features are however difficult to be tested with the current data. We should stress that higher spatial resolution observations are needed to confirm these findings.

\section{Channel maps of CO(1$-$0) emission} \label{appdix:channel}

\begin{figure}[htbp]
\centering
\includegraphics[width=0.8\linewidth]{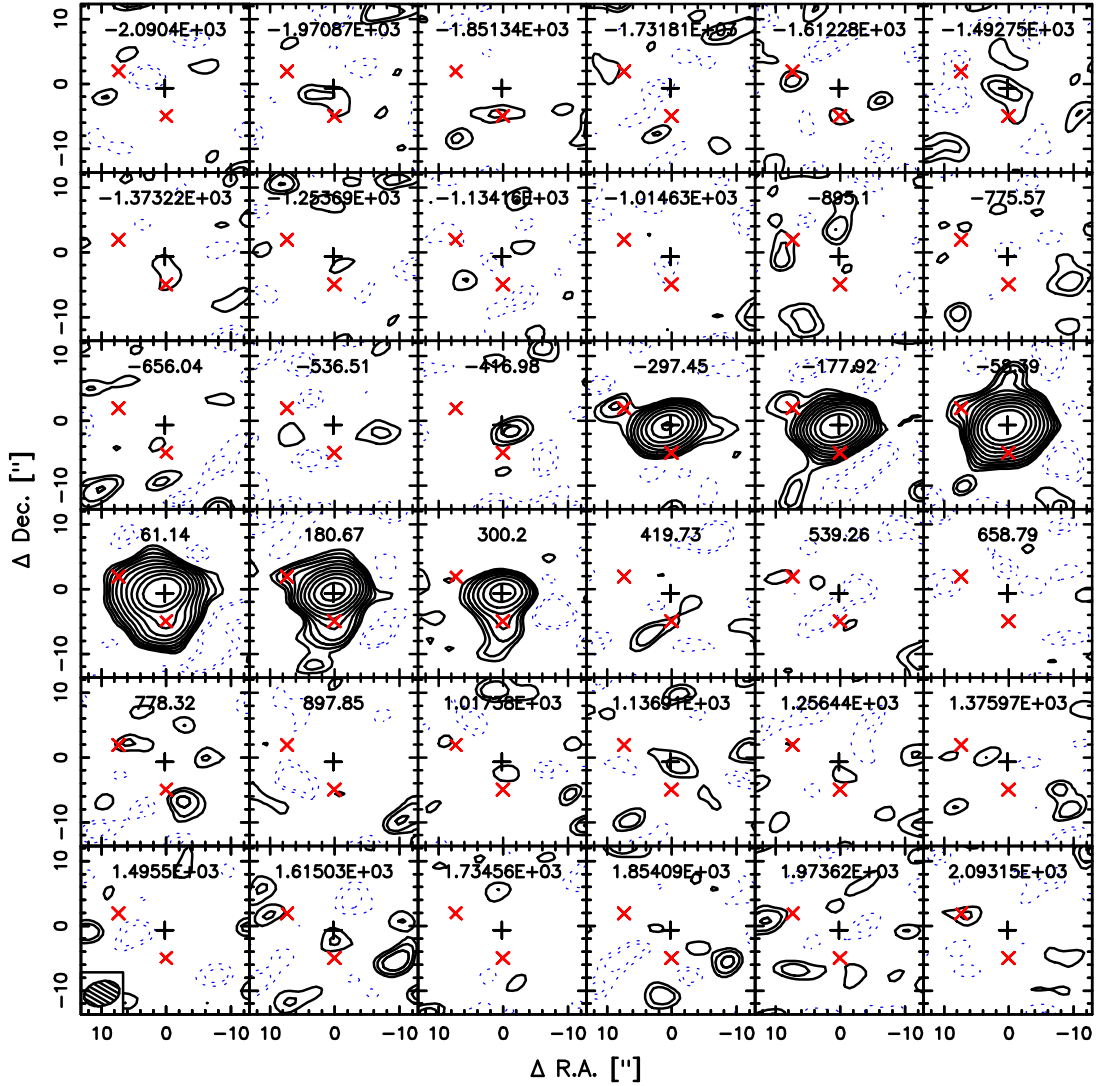}
\caption{Channel maps showing the CO(1$-$0) emission of IRAS~F07599+6508 in the [$-$2000, +1700] \kms\ range with channel wide of $\sim$120 \kms . The red crosses mark the positions of CO subcomponents (F07599S and F07599NE) and the black plus sign indicates the peak of 3~mm continuum emission. Contour levels start at $\pm 1.5 \sigma$ and increase by  factors of 1.5 with positive (negative) contours shown as solid (blue dotted) lines. The velocities in \kms\ are marked in the top left corner of each channel map and the synthesized beam is shown in the bottom-left corner. \label{fig:channel} }
\end{figure}

The spatial and velocity distribution of CO(1$-$0) emission in IRAS~F07599+6508 can be further explored by inspecting the channel maps in Figure~\ref{fig:channel}, which shows the image of CO emission within an interval of $\sim$120 \kms\ encompassing the velocity range from $-$2090 \kms\ to 2093 \kms. The channel maps were produced from the continuum-subtracted data cube that was cleaned down to the 1$\sigma$ noise level. The extended structures to the south and northeast of galaxy center observed in each channel map, spatially coincident with the subcomponent of F07599S and F07599NE (marked as red crosses in Figure~\ref{fig:channel}), show evidence for the reality of such features.

\section{Radio maps and flux measurements}\label{appdix:radio}

Table~\ref{tab:radio} summarizes the radio continuum properties of IRAS~F07599+6508 obtained with VLA. Except the 22.46 GHz data that was obtained in a single CD configuration, observations at the rest of frequencies were made with multiple arrays.

\begin{figure*}[htbp]
\centering
\includegraphics[width=0.24\linewidth]{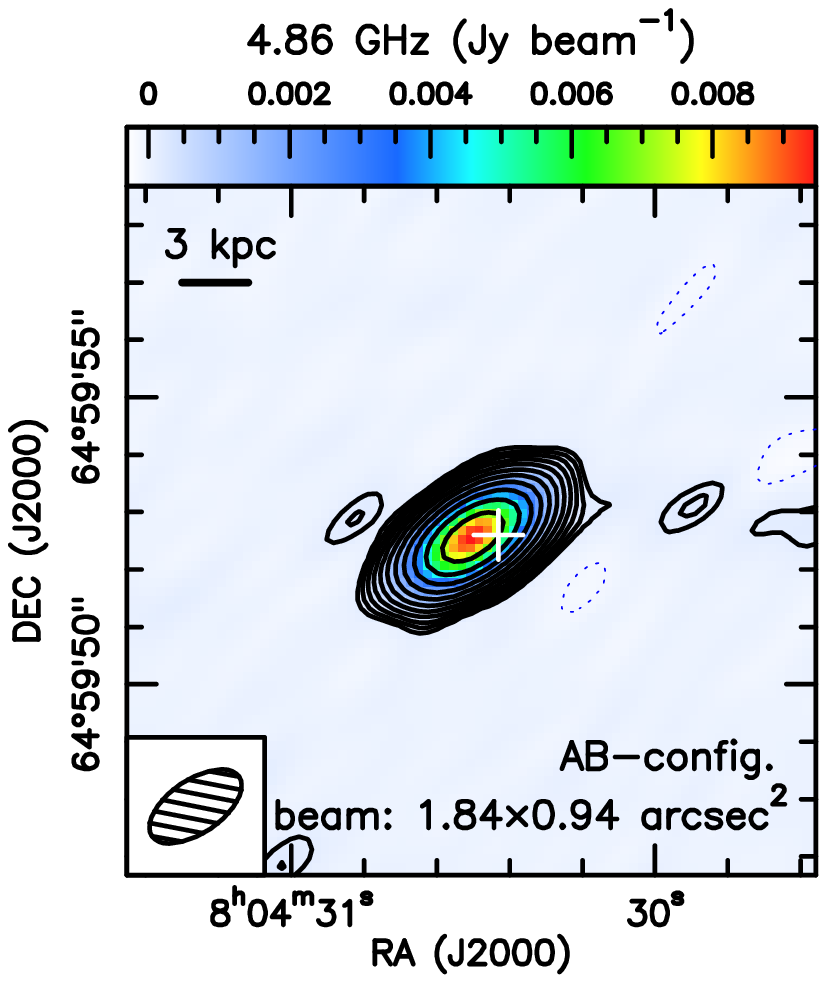}
\includegraphics[width=0.24\linewidth]{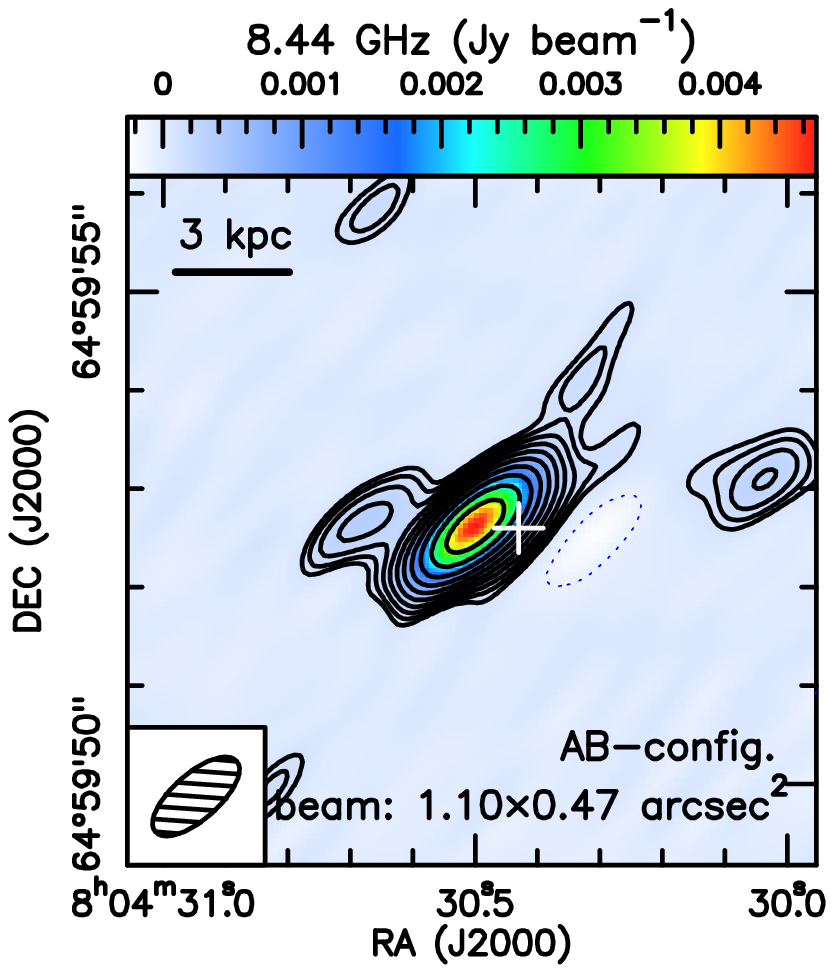}
\includegraphics[width=0.24\linewidth]{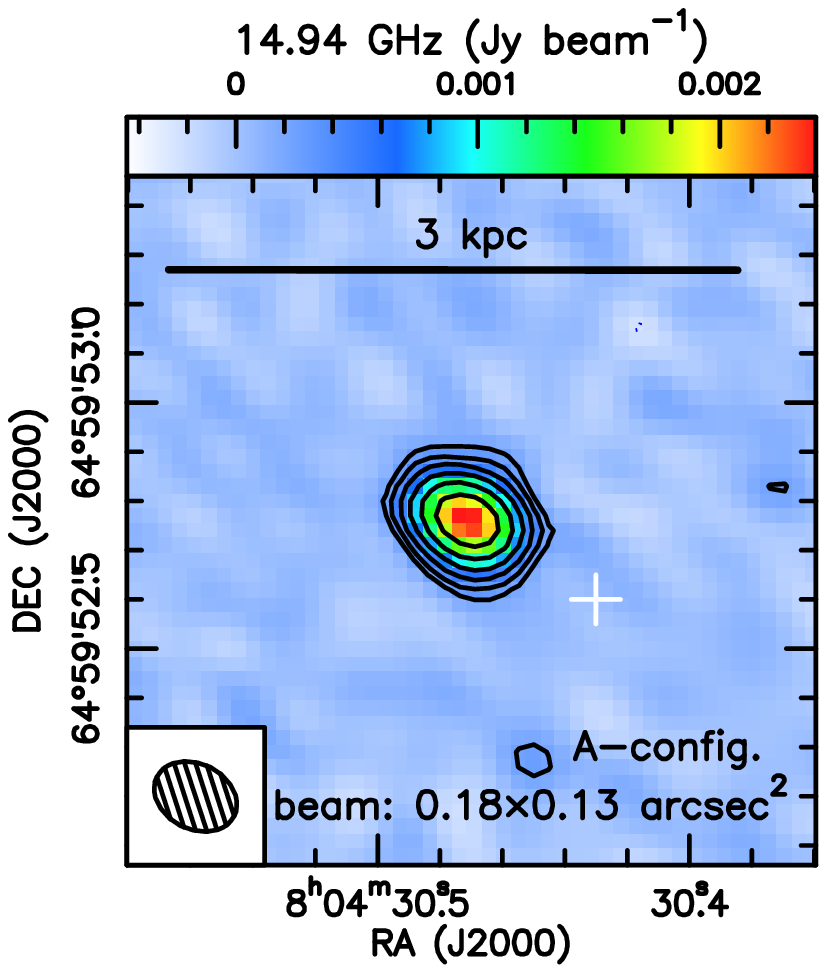}
\includegraphics[width=0.24\linewidth]{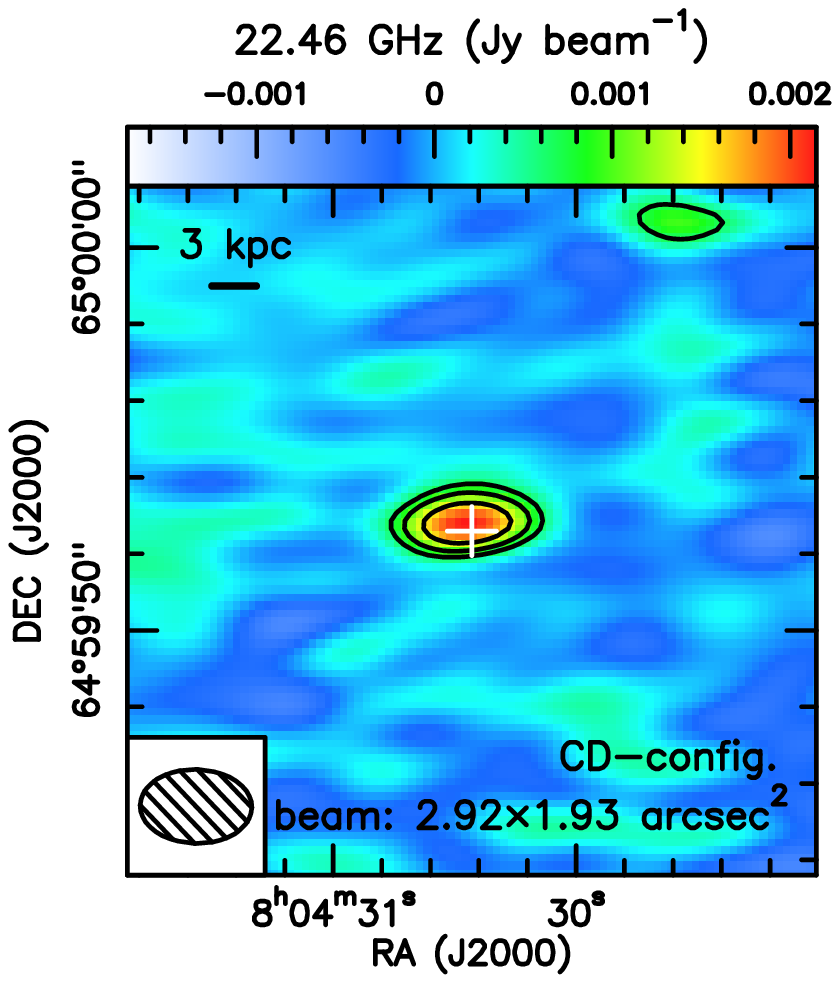}
\caption{VLA multi-band radio continuum imaging of IRAS~F07599+6508 at 4.86, 8.44, 14.94, and 22.46 GHz from left to right. Except the observations at 22.46 GHz which was conducted in a single array of CD configuration, the rest of observations were made with arrays in both AB configuration (A configuration for 14.94 GHz) and CD configuration. Here we only show the maps with higher angular resolution. All the contours start at 3$\sigma$ and increase by a factor of 1.5. The synthesized beam and beam size of each radio observations are shown in the bottom of each panel. The white plus sign indicates the peak of 3~mm continuum emission.\label{fig:radio-map}}
\end{figure*}

To investigate the distribution of radio emission and make a comparison with CO emission, we show the map obtained with higher angular resolution for each frequency in Figure~\ref{fig:radio-map}. For 14.94 and 22.46 GHz, the emission are detected with a compact morphology, while for the emission at 4.86 and 8.44 GHz where the observations were deeper, the radio source exhibits more complex morphology with clumpy structures outside the main body of the galaxy to the east and west, respectively. The peak flux density measured in different configuration were found to be in good agreement for each frequency (i.e., to within $\sim 10\%$; see Table~\ref{tab:radio}). The highest resolution image was obtained at 17.2 GHz with beam size of $\sim$0.\arcsec18 ($\sim$ 0.47 kpc), of which the peak flux density measured is found to be comparable with that measured from CD configuration (beam size $\sim$4.\arcsec58). The consistent peak flux densities measured between different beams for the different bands indicate that the bulk of radio emission probably originates in the inner compact region of the AGN. 

%%%%%%%%%%%%%%%- Fig-10 -%%%%%%%%%%%%%%%%
%%%%%%%%%%%%%%%%%%%%%%%%%%%%%%%%%%%%%%%%

%%%%%%%%%%%%%%%- Fig-11 -%%%%%%%%%%%%%%%%

%%%%%%%%%%%%%%%%%%%%%%%%%%%%%%%%%%%%%%%%

%%%%%%%%%%%%%%- Table-2- %%%%%%%%%%%%%%
\begin{deluxetable}{lllccc}[htbp]
\tablenum{2}
\tabletypesize{\scriptsize}
%\tabletypesize{\small}
%\rotate
%\addtolength{\tabcolsep}{-1.5pt}
\tablecaption{VLA radio continuum properties of IRAS~F07599+6508 \label{tab:radio}}
\tablewidth{0pt}
\tablehead{
\colhead{Obs. freq.} & \colhead{RA\tablenotemark{a}} & \colhead{DEC\tablenotemark{a}} & \colhead{Conf.} & \colhead{Synt.beam} & 
\colhead{S\tablenotemark{b}$_{\rm peak}$} \\
\colhead{(GHz)} & \colhead{(J2000)} & \colhead{(J2000)} & \colhead{} &  \colhead{(arcsec)} &
\colhead{(mJy beam$^{-1}$)}
}
\startdata
1.49 GHz  & 08:04:30.53 & 64:59:53.3 & AB & 5.33$\times$2.08 & 32.33$\pm$0.10 \\
& & & CD & 65.2$\times$28.3 & 36.72$\pm$0.16 \\
4.86 GHz  & 08:04:30.51 & 64:59:52.6 & AB & 1.84$\times$0.94 & 9.59$\pm$0.05 \\
& & & CD & 20.8$\times$8.8 & 9.15$\pm$0.16 \\
8.44 GHz  & 08:04:30.51 & 64:59:52.6 & AB & 1.10$\times$0.47 & 4.67$\pm$0.07 \\
& & & CD & 11.0$\times$5.1 & 5.23$\pm$0.09 \\
14.94 GHz  & 08:04:30.471 & 64:59:52.76 & A & 0.18$\times$0.13 & 2.38$\pm$0.06 \\
& & & CD & 4.58$\times$2.56 & 2.69$\pm$0.08 \\
22.46 GHz  & 08:04:30.46 & 64:59:52.8 & CD & 2.92$\times$1.93 & 2.20$\pm$0.10 \\
\enddata
%% Text for table notes should follow after the \enddata but before
%% the \end{deluxetable}. Make sure there is at least one \tablenotemark
%% in the table for each \tablenotetext.
%\tablecomments{}
\tablenotetext{a}{Source position derived from a 2D Gaussian fit (using \texttt{imfit} task in CASA) to the high-resolution images.}
\tablenotetext{b}{Peak flux density measured from \texttt{imfit} using a single 2D Gaussian component.}
\end{deluxetable}
%%%%%%%%%%%%%%%%%%%%%%%%%%%%%%%%%%%%%%%%

%%%%%%%%%%%%%%%- Fig-12 -%%%%%%%%%%%%%%%%
\begin{figure*}[htbp]
\centering
\includegraphics[width=0.213\linewidth]{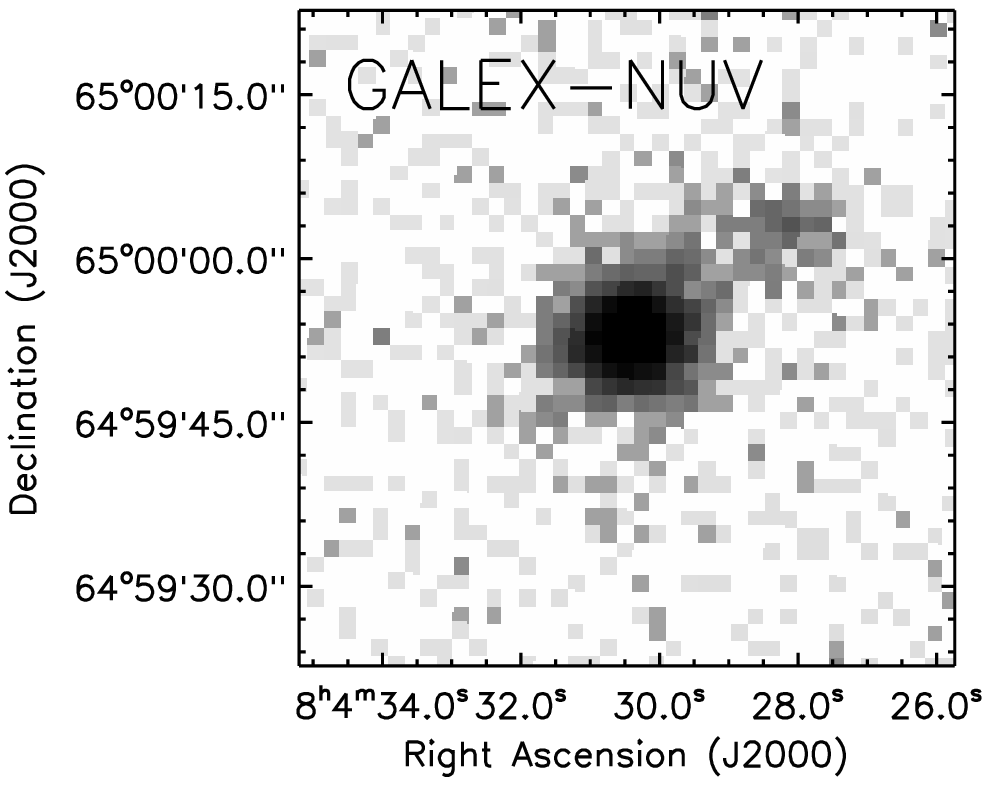}
\hspace{-5pt}
\includegraphics[width=0.15\linewidth]{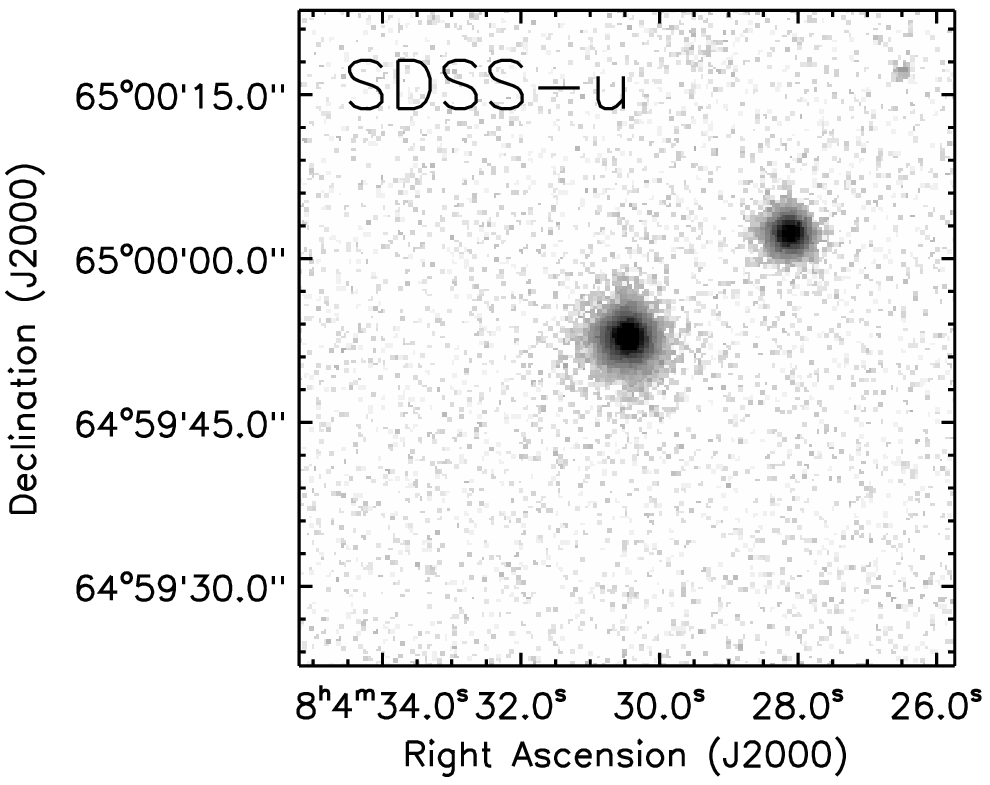}
\hspace{-5pt}
\includegraphics[width=0.15\linewidth]{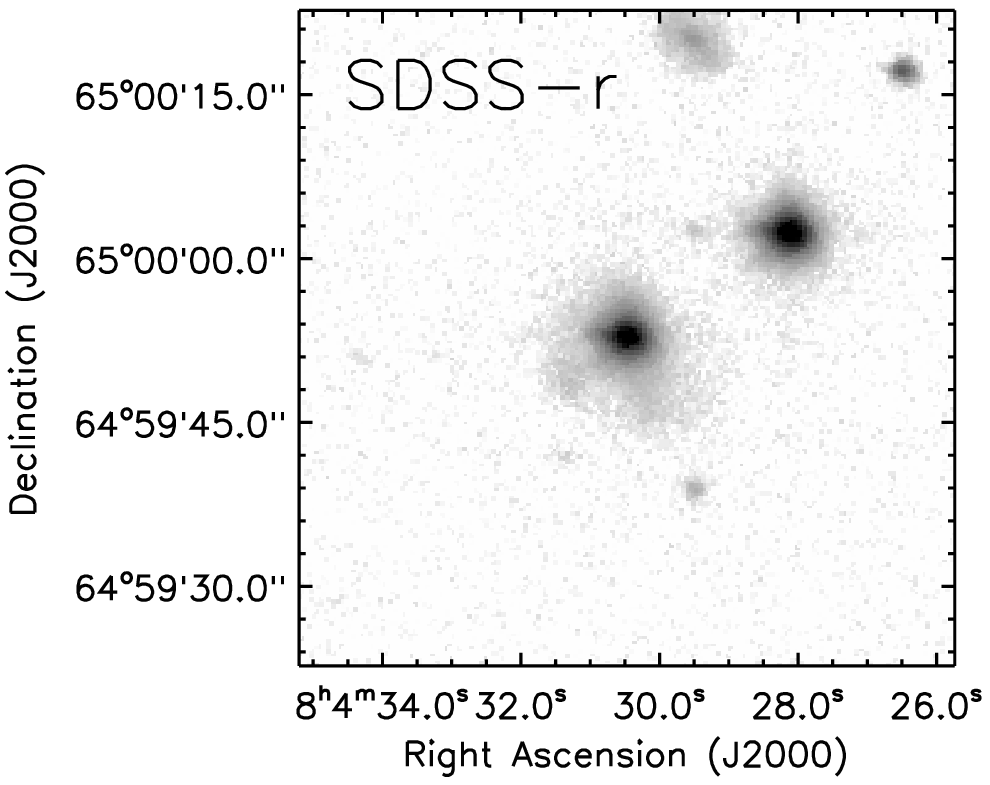}
\hspace{-5pt}
\includegraphics[width=0.15\linewidth]{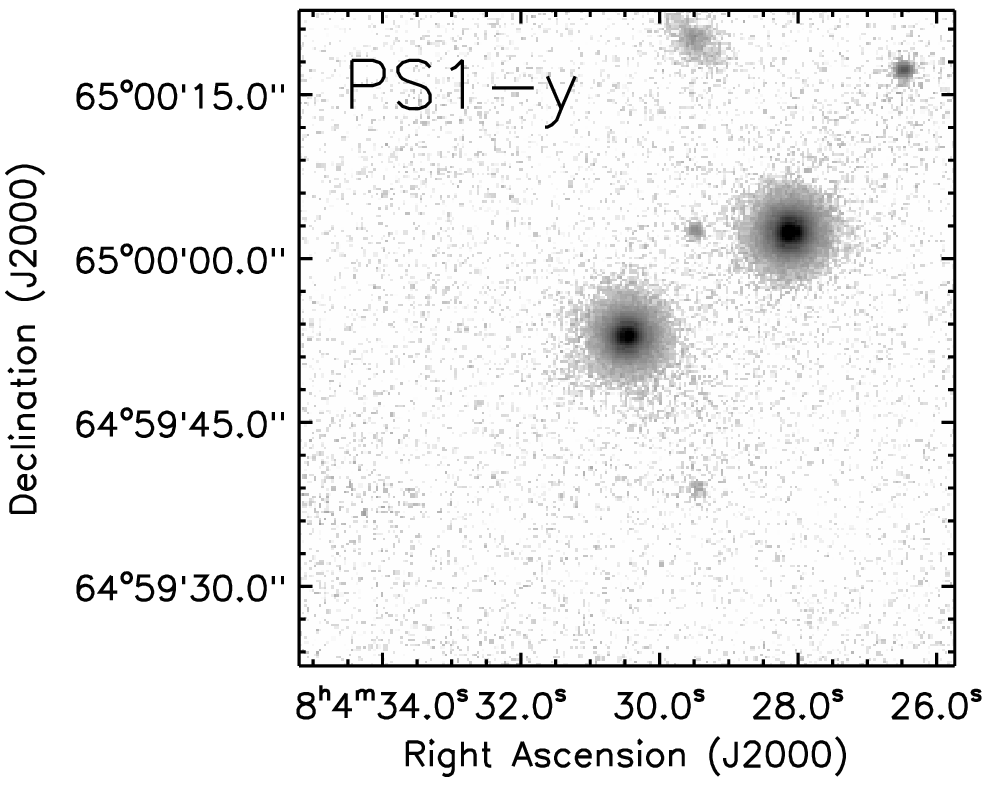}
\hspace{-5pt}
\includegraphics[width=0.15\linewidth]{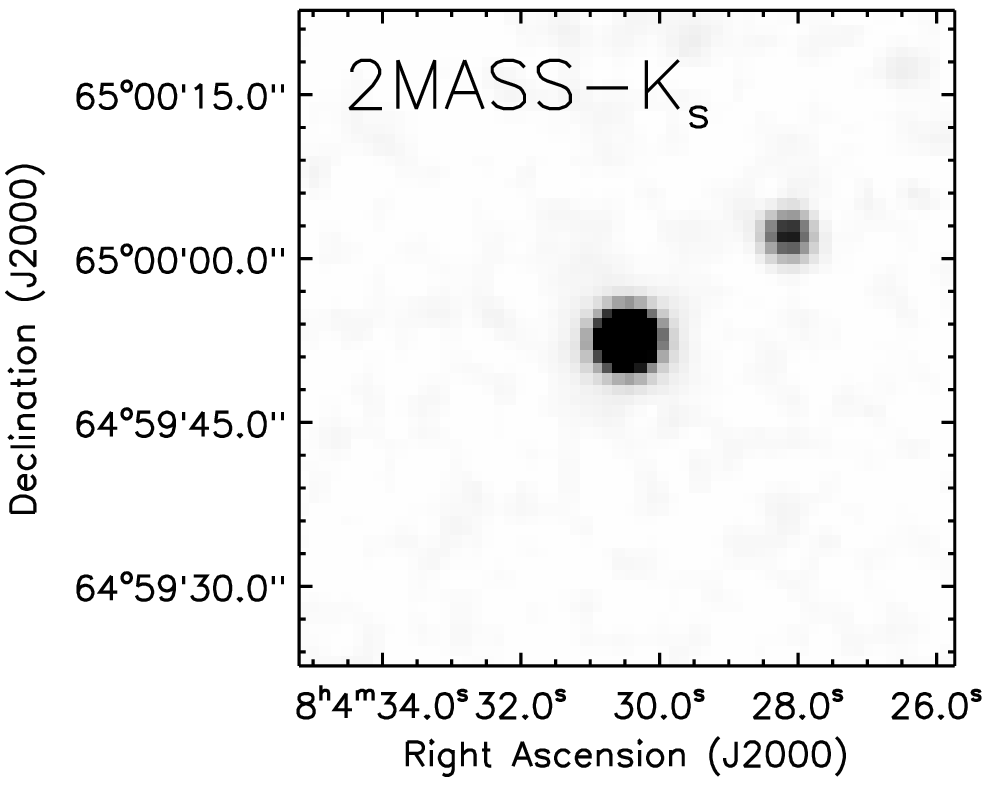}
\hspace{-5pt}
\includegraphics[width=0.15\linewidth]{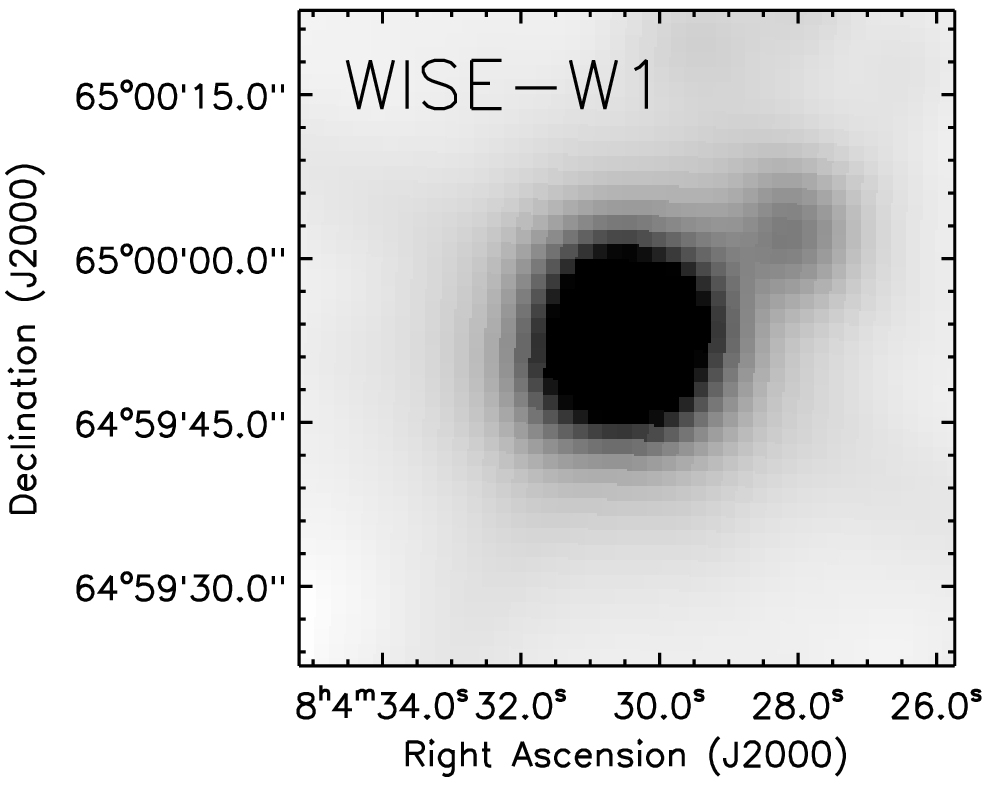}
\caption{Image cutouts (60\arcsec$\times$60\arcsec on a side) of IRAS F07599+6508 and a companion source WISEA J080428.28+650001.7 to the northwest. From left to right, we show the GALEX near-UV, SDSS $u$-, $r$-, Pan-STARRS1 $y$-, 2MASS K$_{\rm s}$-, and WISE 3.4$\mu$m-band images.\label{fig:companion}}
\end{figure*}
%%%%%%%%%%%%%%%%%%%%%%%%%%%%%%%%%%%%%%%%

%%%%%%%%%%%%%%%- Fig-13 -%%%%%%%%%%%%%%%%
\begin{figure*}[htbp]
\centering
\includegraphics[width=0.85\linewidth]{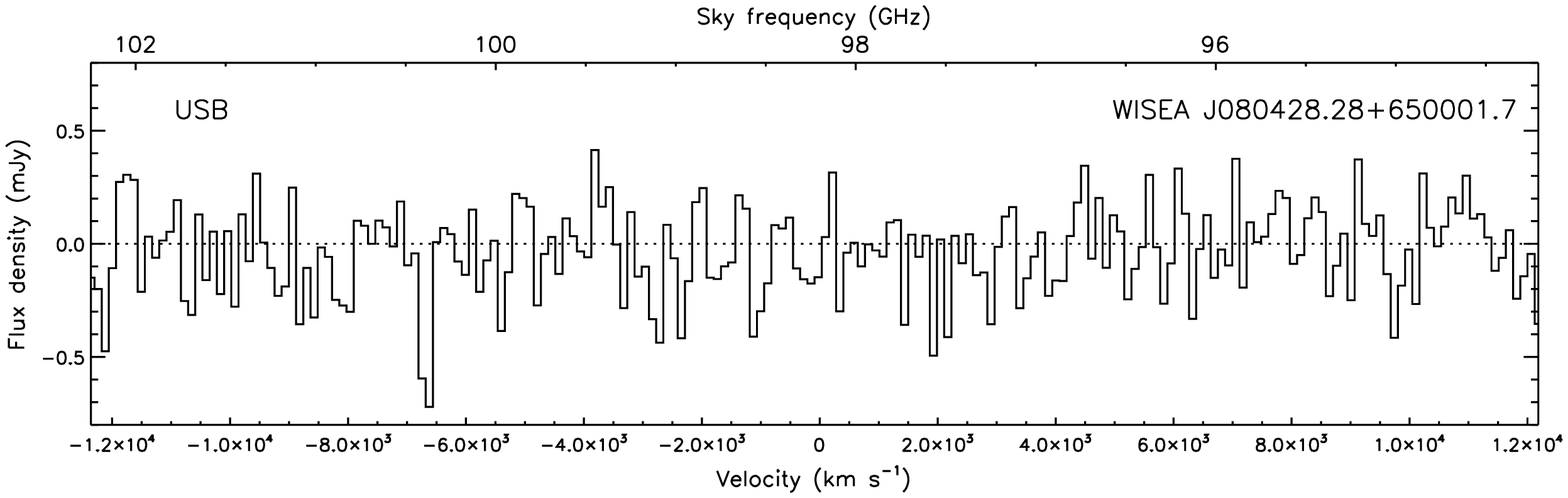}
\includegraphics[width=0.85\linewidth]{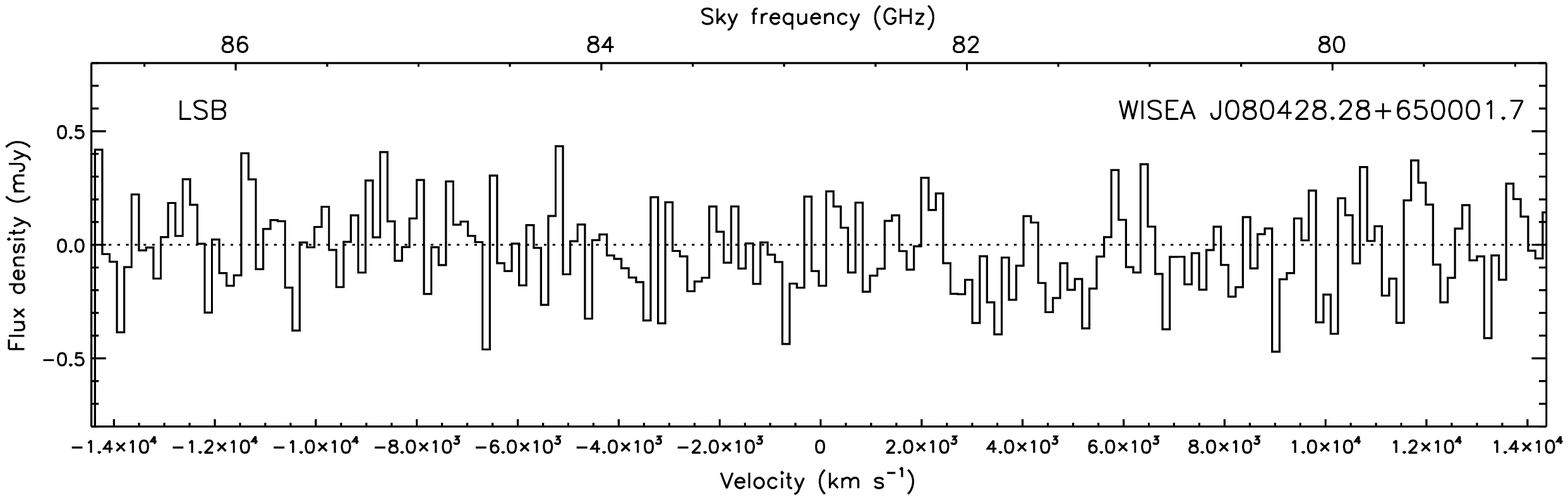}
\caption{NOEMA spectrum (without correction for PBA) of WISEA J080428.28+650001.7 covered in the USB (top) and the LSB (bottom). The spectra are extracted by fitting a point-source model to the $uv$-data and are binned to a channel width of $\sim120-140$ \kms.\label{fig:comp-spect}}
\end{figure*}
%%%%%%%%%%%%%%%%%%%%%%%%%%%%%%%%%%%%%%%%

\section{Multi-wavelength data of IRAS~F07599+6508 and a companion source to the northwest}\label{appdix:multidata}

In this appendix we show the multi-wavelength (from GALEX near-UV to WISE 3.4$\mu$m band) data of IRAS F07599+6508 and a companion source WISEA J080428.28+650001.7 to the northwest (Figure~\ref{fig:companion}). J0804+6500 has a distance of 17.\arcsec5 ($\sim$ 45 kpc) from IRAS F07599+6508, with a coordinates of 08:04:28.140, +65:00:02.25.

Figure~\ref{fig:comp-spect} shows the NOEMA spectra (without correction for primary beam attenuation (PBA)) extracted from the position of J0804+6500 by fitting a point-source model to the $uv$-data. The tuning frequency covered in the upper and lower sidebands corresponds to a redshift range of 0.128 to 0.223 and 0.238 to 0.462, respectively, if associated with a CO(1$-$0) transition. The 1$\sigma$ rms noise level of the spectrum in the USB and LSB shown in Figure~\ref{fig:comp-spect} is 0.20 mJy and 0.21 mJy, respectively. We do not detect any significant emission from J0804+6500 in our observations.


\bibliography{reference}{}
\bibliographystyle{aasjournal}



\end{document}